\documentclass[reprint, prmaterials, amsfonts, amssymb, amsmath, aps, floatfix, article]{revtex4-2}
\usepackage[pdfusetitle]{hyperref}
\usepackage{graphicx}
\usepackage{dcolumn}
\newcolumntype{d}[1]{D{.}{.}{#1}}
\usepackage{bm}
\usepackage{longtable}
\usepackage{mathtools}

\usepackage{xspace}
\usepackage{xpunctuate}
\DeclareRobustCommand\etal{\xperiodafter{\emph{et al}}}

\DeclareRobustCommand\fu{\xperiodafter{{f.u}}}

\hypersetup{
  colorlinks=true,
  linkcolor=blue,
  filecolor=blue,
  citecolor=blue,
  urlcolor=blue,
}

\newcommand{\svs}{ScV$_6$Sn$_6$\xspace}

\usepackage{float}

\begin{document}

\title{Order-by-disorder charge density wave condensation at $\mathbf{
    \textit{q} =(\frac{1}{3},\frac{1}{3},\frac{1}{3})}$ in kagome
  metal ScV$_6$Sn$_6$}



\author{Alaska Subedi} 

\affiliation{CPHT, CNRS, \'Ecole polytechnique, Institut Polytechnique
  de Paris, 91120 Palaiseau, France}

\date{\today}

\begin{abstract}
  The recent discovery of a charge density wave order at the wave
  vector $P$ $(\frac{1}{3},\frac{1}{3},\frac{1}{3})$ in the kagome
  metal ScV$_6$Sn$_6$ has created a mystery because subsequent
  theoretical and experimental studies show a dominant phonon
  instability instead at another wave vector $H$
  $(\frac{1}{3},\frac{1}{3},\frac{1}{2})$.  In this paper, I use first
  principles total energy calculations to map out the landscape of the
  structural distortions due to the unstable phonon modes at $H$, $L$
  $(\frac{1}{2},0,\frac{1}{2})$, and $P$ present in this material.  In
  agreement with previous results, I find that the distortions due to
  the $H$ instability cause the largest gain in energy relative to the
  parent structure, followed in order by the $L$ and $P$
  instabilities.  However, only two distinct structure occur due to
  this instability, which are separated by 6 meV/f.u.  The instability
  at $L$ results in three distinct structures separated in energy by 5
  meV/f.u.  In contrast, six different distorted structures are
  stabilized due to the instability at $P$, and they all lie within 2
  meV/f.u.\ of each other.  Hence, despite a lower energy gain, the
  condensation at $P$ could be favorable due to a larger entropy gain
  associated with the fluctuations within a manifold  with larger
  multiplicity via the order-by-disorder mechanism.  
\end{abstract}

\maketitle

\section{Introduction}

Materials that have the kagome lattice as their structural motif have
been well studied in the context of frustrated magnetism
\cite{Mendels2016}.  The frustrated lattice also gives rise to
remarkable electronic structure with Dirac cones and flat bands
\cite{Beugeling2012,Meier2020,Kang2019,Liu2020}.  Various emergent
phases due to electronic instabilities in these materials have been
anticipated
\cite{Guo2009,Ruegg2011,Tang2011,Wang2013,Kiesel2013,Mazin2014}, and
the discovery of charge density wave (CDW) order in kagome metals
$A$V$_3$Sb$_5$ ($A$ = K, Rb, Cs) and FeGe has motivated further
exploration of this class of materials for uncommon ground states and
excitations \cite{Ortiz2019,Teng2022}.

A notable result of this activity is Suriya Arachchige \etal's
discovery of first-order CDW transition in the bilayer kagome metal
ScV$_6$Sn$_6$ at the wave vector $P$
$(\frac{1}{3},\frac{1}{3},\frac{1}{3})$ \cite{Arachchige2022}, which
is highly unusual because $P$ is not a high-symmetry point in the
Brillouin zone where dense scattering phase space is usually expected.
Angle resolved photoemission and optical spectroscopy experiments
across the transition show relatively modest electronic structure
changes at the Fermi level with no gap opening
\cite{Tuniz2023,Cheng2023,Kang2023,Hu2023a,Hu2023b,Korshunov2023,
  Lee2023,DiSante2023}, which indicates that an electronic instability
does not cause the CDW transition.
Intriguingly, inelastic x-ray scattering (IXS) experiments find that
the lowest-frequency phonon mode at $P$ softens only modestly as the
transition temperature $T_c = 92$ K is approached from above
\cite{Cao2023,Korshunov2023}, while a phonon mode at another wave
vector $H$ $(\frac{1}{3},\frac{1}{3},\frac{1}{2})$ softens completely
to zero \cite{Korshunov2023}. Diffuse scattering signals that grow in
intensity from room temperature to 100 K is observed around $H$,
whereas no such signals are seen around the ordering wave vector $P$
\cite{Cao2023,Korshunov2023,Pokharel2023}.  Calculated phonon dispersion find an
unstable branch that has the largest imaginary frequency at $H$
\cite{Tan2023}, seemingly in agreement with the IXS experiments.
However, the calculated energy gain due to the instability at $H$ is also
larger than that due to the one at $P$.  This suggests that a mechanism
beyond the harmonic level in atomic displacements may be necessary to
describe the CDW transition observed in this material, 
%
although the
energetics of all possible distorted structures due to these
phonon instabilities has not been thoroughly investigated.  

In this paper, I study the energetics of structural distortions in
ScV$_6$Sn$_6$ using group theory and density functional theory (DFT)
calculations.  Group-theoretical analysis was utilized to enumerate
the possible symmetrically-distinct distortions in the order parameter
subspaces described by the phonon instabilities at $H$, $L$
$(\frac{1}{2},0,\frac{1}{2})$, and $P$.  Calculated eigenvectors of
these phonon instabilities were then used to generate the distorted
structures, and their total energies were obtained from DFT after full
relaxations that minimized both the lattice stresses and atomic
forces.  Consistent with previous results, I find that a distorted
structure due to the instability at $H$ has the lowest energy.  I was
also able to stabilize six symmetrically-distinct structures due to
the instability at $P$ whose calculated total energies lie within 2
meV/\fu of each other.  Fluctuations among these nearly-degenerate
states likely stabilize the CDW order at $P$ via the order-by-disorder
mechanism.  Experimental confirmation of the presence of these
nearly-degenerate states by, for example, counting the number of
domains in the low-temperature phase would strengthen the case that
entropic force is responsible for the structural transition observed in
this material.
%

\section{Computational Approach}

The phonon dispersions and structural relaxation calculations
presented here were performed using the pseuodopotential-based {\sc
  quantum espresso} package \cite{qe} within the optB88-vdW
approximation for the exchange-correlation functional \cite{optb88}.
I used the pseudopotentials generated by Dal Corso \cite{pslib} and
plane-wave cutoffs of 60 and 600 Ry for the basis-set and charge
density expansions, respectively.  The Brillouin zone integration was
performed using a $12\times12\times6$ $k$-point grid for the parent
13-atom structure.  Equivalent or denser grids were used for the
calculations on supercells. A $0.01$ Ry Marzari-Vanderbilt smearing
was used to determine the partial occupancies. The dynamical matrices
of the parent structure were calculated on a $6\times6\times6$
$q$-point grid using density functional perturbation theory
\cite{dfpt}. 

I used the {\sc isotropy} code \cite{isotropy} to determine the order
parameter directions of all possible distortions due to the unstable
phonon modes at $H$, $L$ and $P$, as well as the number of domains
exhibited by the distorted structures.  The calculated phonon
eigenvectors of the unstable modes were used to generate the distorted
structures corresponding to the isotropy subgroups on the supercells
commensurate with the phonon wave vectors, which were then fully
relaxed by minimizing both the lattice stresses and atomic forces. I
made extensive use of the {\sc findsym} \cite{findsym} and {\sc spglib}
\cite{spglib} codes in the symmetry analysis of the calculated
structures.  The {\sc amplimodes} code \cite{ampli} was used to determine the
order parameter amplitudes of the relaxed structure.  The polynomial
expansion of the free energy as a function of the order parameters
were performed using the {\sc invariants} code \cite{inv}.

\section{Results and Discussion}

\begin{figure}[!htbp]
  \includegraphics[width=\columnwidth]{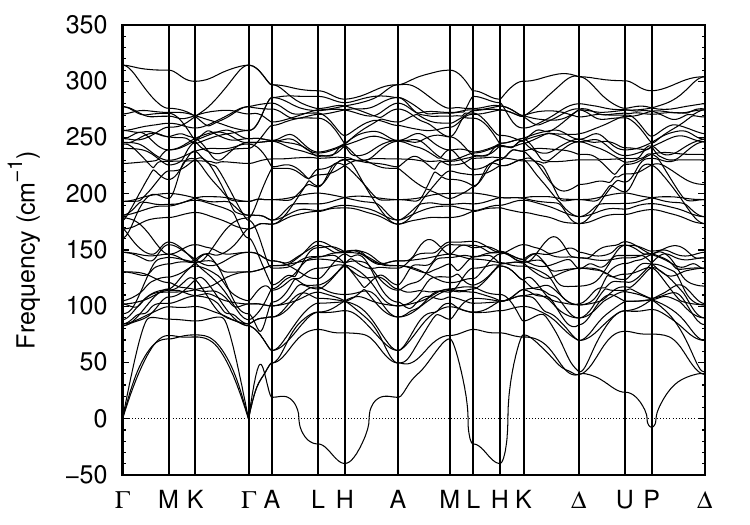}
  \caption{Calculated phonon dispersions of fully-relaxed
    ScV$_6$Sn$_6$ in the parent $P6/mmm$ phase obtained using the
    optB88-vdW functional.  The high-symmetry points are $\Gamma$
    $(0,0,0)$, $M$ $(\frac{1}{2},0,0)$, $K$ $(\frac{1}{3},
    \frac{1}{3}, 0)$, $A$ $(0,0,\frac{1}{2})$, $L$ $(\frac{1}{2}, 0,
    \frac{1}{2})$, $H$ $(\frac{1}{3}, \frac{1}{3}, \frac{1}{2})$,
    $\Delta$ $(0,0,\frac{1}{3})$, $U$ $(\frac{1}{2},0,\frac{1}{3})$,
    $P$ $(\frac{1}{3}, \frac{1}{3}, \frac{1}{3})$. Imaginary
    frequencies are denoted by negative values.}
  \label{fig:ph}
\end{figure}

The calculated phonon dispersions of \svs in the high-temperature
$P6/mmm$ structure obtained using the optB88-vdW functional is shown
in Fig.~1.  As in previously published results
\cite{Tan2023,Hu2023b,Cao2023,Korshunov2023,Lee2023}, there is a
nondegenerate branch that is unstable along the path $L$-$H$ and at
$P$. The largest instability occurs at $H$ with a calculated imaginary
frequency of 40$i$ cm$^{-1}$.  For comparison, the calculated values
at $L$ and $P$ are 22$i$ and 7$i$ cm$^{-1}$, respectively.  These
results are somewhat in agreement with the inelastic x-ray scattering
experiment of Korshunov \etal in that they observe a complete
softening of the phonon mode at $H$ as $T_c$ is approached from above,
whereas the phonons at $L$ and $P$ exhibit only a weak softening
\cite{Korshunov2023}.  However, these are incongruent with the fact
that the CDW order stabilizes at $P$, where the calculated instability
is the weakest.  The gain in energy due to the structural distortion
at $P$ could in principle be larger than the one at $H$ as a result of
additional freezing of secondary order parameters.  But full
structural relaxations by Tan and Yan show that the distorted
structure due to the instability at $H$ is lower in energy than the
one due to the instability at $P$ \cite{Tan2023}.

\begin{table}[!htbp]
    \caption{\label{tab:iso} Isotropy subgroups and order parameter
      directions (OPD) of $P6/mmm$ for the irreps $H_3^{}$, $L_2^-$,
      and $P_1^{}$.  Total energies of the fully-relaxed structures
      corresponding to these order parameters are given in the units
      of meV per formula unit relative to the parent $P6/mmm$ phase.
      Note that the order parameters refer to that of the initial
      structure before the minimization of stresses and forces.
      Symmetry-allowed secondary order parameters appear for the final
      relaxed structures. Not all distortions could be
      stabilized. Full structural information of all the phases that
      could be stabilized is given in Supplemental Material~\cite{sm}.}
       
    \begin{ruledtabular}
      
      \begin{tabular}{l l d{3.2}}
        Space group (No.)  & OPD & \multicolumn{1}{c}{Energy (meV/\fu)} \\
        \hline
        $P6/mmm$ (191) &  $H_3^{}(a,0)$ &        -16.68 \\
        $P6_3/mmc$ (194) & $H_3^{}(0,a)$ &       -10.64 \\
        $P\overline{6}m2$ (187) &  $H_3^{}(a,b)$ & \multicolumn{1}{c}{---} \\
        $Immm$ (71) & $L_2^-(a,0,0)$ & -12.10 \\
        $Fmmm$ (69) & $L_2^-(a,-a,0)$ & -4.01 \\
        $P6/mmm$ (191) & $L_2^-(a,a,a)$ & -7.50 \\
        $C2/m$ (12) & $L_2^-(a,b,0)$ & \multicolumn{1}{c}{---} \\
        $Cmmm$ (65) & $L_2^-(a,b,a)$ & \multicolumn{1}{c}{---} \\
        $P2/m$ (10) & $L_2^-(a,b,c)$ & \multicolumn{1}{c}{---} \\
        $R\overline{3}m$ (166) & $P_1^{}(a,0,0,0)$ & -3.07 \\ 
        $R\overline{3}m$ (166) & $P_1^{}(-a,0,0,0)$ & -1.17 \\
        $P6/mmm$ (191) & $P_1^{}(a,0,a,0)$ & -2.18 \\
        $P6/mmm$ (191) & $P_1^{}(-a,0,-a,0)$ & -1.76 \\
        $P6mm$ (183) & $P_1^{}(a,b,a,b)$ & -2.20 \\
        $P6mm$ (183) & $P_1^{}(-a,-b,-a,-b)$ & -1.92 \\
        $R3m$ (160) & $P_1^{}(a,b,0,0)$ & \multicolumn{1}{c}{---} \\
        $P\overline{6}m2$ (187) & $P_1^{}(a,b,a,-b)$ & \multicolumn{1}{c}{---} \\
        $P\overline{3}m1$ (164) & $P_1^{}(a,0,b,0)$ & \multicolumn{1}{c}{---} \\
        $P3m1$ (156) & $P_1^{}(a,b,c,d)$ & \multicolumn{1}{c}{---} \\
      \end{tabular}
    \end{ruledtabular}
    
\end{table}

Previous structural relaxation studies report only one distorted structure due
to the instability at $P$ \cite{Tan2023,Cao2023,Gu2023}, and it is possible that another distorted structure at
$P$ lies at the global minimum in the energy landscape. In fact, even
though the unstable phonon branch is nondegenerate, the order
parameter subspaces due to the instabilities at $H$, $L$, and $P$ are
multidimensional because the stars of these points have two, three,
and four elements, respectively.  The corresponding stars are $H$
$\{(\frac{1}{3},\frac{1}{3},\frac{1}{2}),
(\frac{2}{3},\frac{2}{3},\frac{1}{2})\}$, $L$
$\{(\frac{1}{2},0,\frac{1}{2}), (0,\frac{1}{2},\frac{1}{2}),
(\frac{1}{2}, \frac{1}{2}, \frac{1}{2})\}$, and $P$
$\{(\frac{1}{3},\frac{1}{3},\frac{1}{3}),
(\frac{2}{3},\frac{2}{3},\frac{1}{3}),
(\frac{1}{3},\frac{1}{3},\frac{2}{3}),
(\frac{2}{3},\frac{2}{3},\frac{2}{3})\}$.  The elements of a star
correspond to distinct directions that span the subspace of all atomic
displacements generated by the eigenvector of the unstable phonon
mode.  The phonon instabilities at $H$, $L$, and $P$ have the
irreducible representations (irreps) $H_3^{}$, $L_2^-$, and $P_1^{}$,
respectively.  The isotropy subgroups of an irrep enumerate all
possible low-symmetry space groups that can arise out of the
corresponding phonon instability, and Table~\ref{tab:iso} lists the
isotropy subgroups and the associated order parameter directions due
to these instabilities.  I used the calculated eigenvectors of the
unstable phonon modes to generate all these structures.  Thus
generated structures were then fully relaxed by minimizing both
lattice stresses and atomic forces.  Relaxation takes a structure to
the local minimum in the manifold of the energy landscape defined by
the respective order parameter direction.

The calculated total energies of the distorted structures
corresponding to the isotropy subgroups obtained after strucutral
relaxations are also given in Table~\ref{tab:iso} relative to the
energy of the parent $P6/mmm$ phase.  In agreement with the finding of
Tan and Yan \cite{Tan2023}, the $H_3(a,0)$ distortion which also
belongs to the space group $P6/mmm$ has the lowest energy.  However, I
was able to stabilize several more distorted structures, although not
all possible isotropy subgroups due to the $H_3^{}$, $L_2^-$, and
$P_1^{}$ instabilities could be stabilized during structural
relaxation.  The unstable order parameters presumably lie at local
maxima or saddle points in the energy landscape, and they relaxed to
one of the higher-symmetry phases during the relaxation process.

There are three isotropy subgroups of the $H_3^{}$ irrep, out of which
the $P6/mmm$ $H_3^{}(a,0)$ and $P6_3/mmc$ $H_3^{}(0,a)$ structures
could be stabilized with energies of $-$16.68 and $-$10.64 meV/f.u.,
respectively, relative to that of the parent structure.  Six isotropy
subgroups belong to the $L_2^-$ irrep, but only three of them
maintained their symmetry during the structural relaxation. They are
$Immm$ $L_2^-(a,0,0)$, $Fmmm$ $L_2^-(a,a,a)$, and $P6/mmm$
$L_2^-(a,-a,0)$ with relative energies of $-$12.10, $-$4.01, and
$-$7.50 meV/f.u., respectively.

\begin{figure}[!htbp]
  \includegraphics[width=\columnwidth]{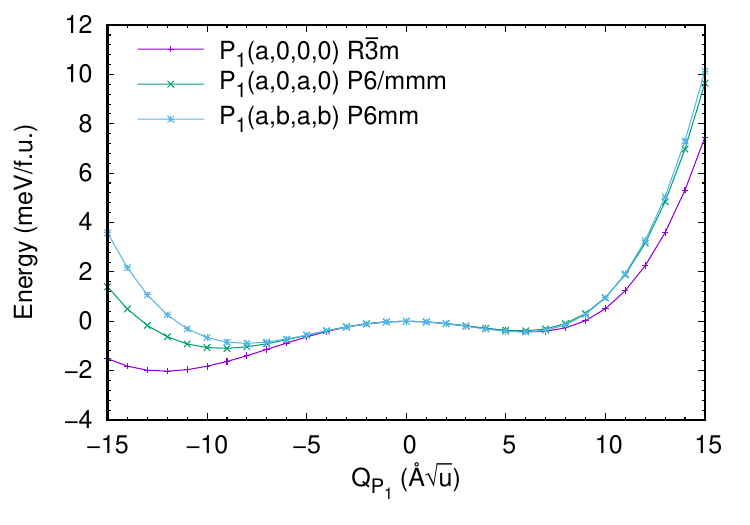}
  \caption{Calculated energy curves of ScV$_6$Sn$_6$ along three
    different order parameter directions due to the $P_1^{}$
    instability illustrating the presence of six
    symmetrically-distinct local minima.  The energy curves are
    asymmetric due to the presence of odd-order terms in the free
    energy. For $P_1^{}(a,b,a,b)$, $b = a \cos(75^\circ)$ has been
    used that gives the deepest local minimum. Note that the total
    energies of the structures at the local minima are further lowered
    and arrive at the values given in Table~\ref{tab:iso} after
    structural relaxations. }
  \label{fig:curves}
\end{figure}

\begin{table*}[!htbp]
    \caption{\label{tab:opdcomp} Mode amplitudes (\AA) of the primary
      and secondary order parameter directions for the six distorted
      structures that could be stabilized due to the $P_1^{}$ phonon
      instability.  $\Gamma_1^+$ and $K_1$ are one- and
      two-dimensional irreps, respectively.  The OPD of $K_1$ is
      $(a,0)$ for all structures.  $d_{\textrm{Sc-Sn}}^1$,
      $d_{\textrm{Sc-Sn}}^2$, and $d_{\textrm{Sc-Sn}}^3$ denote the
      three shortest nearest-neighbor Sc-Sn distances in \AA\ in
      increasing order. }
       
    \begin{ruledtabular}
      
      \begin{tabular}{l l  d{1.4} d{1.4} c d{1.4} d{1.4} d{1.4} d{1.4} d{1.4}}
        Space group &   \multicolumn{2}{c}{$P_1^{}$} &
        \multicolumn{1}{c}{$\Gamma_1^+$} &
        \multicolumn{2}{c}{$\Delta_1^{}$} &
        \multicolumn{1}{c}{$K_1^{}$} & 
        \multicolumn{1}{c}{$d_{\textrm{Sc-Sn}}^1$} &
        \multicolumn{1}{c}{$d_{\textrm{Sc-Sn}}^2$} &
        \multicolumn{1}{c}{$d_{\textrm{Sc-Sn}}^3$} 
        \\
         &  
        OPD & \multicolumn{1}{c}{amplitude} & 
        \multicolumn{1}{c}{amplitude} & 
        OPD & \multicolumn{1}{c}{amplitude} & 
        \multicolumn{1}{c}{amplitude}  \\
        \hline
        $R\overline{3}m$  & $(a,0,0,0)$     &  0.1843 &  0.0024 & ---     &  \multicolumn{1}{c}{---} &  \multicolumn{1}{c}{---}  & 2.9108 & 2.9323 & 2.9722 \\
        $R\overline{3}m$  & $(-a,0,0,0)$    &  0.0981 &  0.0040 & ---     &  \multicolumn{1}{c}{---} &  \multicolumn{1}{c}{---} & 2.9161 & 2.9368 & 2.9501 \\
        $P6/mmm$          & $(a,0,a,0)$     &  0.1477 &  0.0028 & $(a,0)$ &  0.0066 &   0.0060 & 2.9109 & 2.9125 & 2.9334 \\
        $P6/mmm$          & $(-a,0,-a,0)$   &  0.1287 &  0.0039 & $(a,0)$ &  0.0412 &   0.0028 & 2.9048 & 2.9159 & 2.9330 \\
        $P6mm$            & $(a,b,a,b)$     &  0.1463 &  0.0028 & $(a,b)$ &  0.0094 &   0.0058 & 2.9103 & 2.9120 & 2.9122 \\
        $P6mm$            & $(-a,-b,-a,-b)$ &  0.1299 &  0.0038 & $(a,b)$ &  0.0389 &   0.0026 & 2.9032 & 2.9065 & 2.9151 \\
      \end{tabular}
    \end{ruledtabular}
    
\end{table*}

The $P_1^{}$ irrep has seven isotropy subgroups, but only structures
belonging to the $R\overline{3}m$ $P_1^{}(a,0,0,0)$, $P6/mmm$
$P_1^{}(a,0,a,0)$, and $P6mm$ $P_1^{}(a,b,a,b)$ subgroups remained
stable during relaxation.  Interestingly, for each of these three
subgroups, two distinct structures could be stabilized that are
characterized by order parameters that are out-of-phase by
180$^{\circ}$.  The out-of-phase pairs occur at different magnitudes
of the order parameter, which indicates the presence of local minima
at asymmetric positions in the energy landscape defined by the order
parameter subspace.  This can happen because odd-order terms are
allowed by symmetry in the polynomial expansion of the free energy as
a function of the order parameters associated with the $P_1^{}$ irrep,
while only even-order terms are allowed for those of $H_3^{}$ and
$L_2^-$ irreps.  Nevertheless, the presence of odd-order terms only
guarantees that energy surface be asymmetric with respect to the order
parameter direction and does not necessitate multiple minima.
Therefore, the occurrence three out-of-phase pairs at different values
of the order parameters in the energy landscape is noteworthy.
Fig.~\ref{fig:curves} shows the energy curves along the order
parameter directions $P_1^{}(a,0,0,0)$, $P_1^{}(a,0,a,0)$, and
$P_1^{}(a,b,a,b)$ that illustrates the presence of six local minima in
this manifold.

As one can note from Table~\ref{tab:iso}, the calculated total
energies of the six distorted phases due to the $P_1^{}$ instability
lies within 2 meV/\fu of each other, with values ranging from $-$3.07 to
$-$1.17 meV/\fu relative to that of the parent phase.  Despite this
near-degeneracy, these structures can be distinguished by the
different values of mode amplitudes for the primary and secondary
order parameter directions, which are given in
Table~\ref{tab:opdcomp}. For example, the two $R\overline{3}m$ phases
with the $P_1^{}(a,0,0,0)$ order parameter direction out-of-phase
by 180$^\circ$ have amplitudes of 0.1843 and 0.0981 \AA\ for this mode. 
There is only one nearest-neighbor Sc-Sn distance in the out-of-plane
direction in the parent structure, which multiplies into different values
in the distorted structures.  The three shortest nearest-neighbor
distances in these six structures are also given in
Table~\ref{tab:opdcomp}, and they can also be used to distinguish
these structures.

Although the calculated energies of the distorted phases due to the
$P_1^{}$ instability are higher relative to those due to the $H_3^{}$
and $L_2^-$ instabilities, more nearly-degenerate distinct structures
occur in the manifold of the energy landscape generated by $P_1^{}$.
As a result, there is more phase space for fluctuations in the order
parameter subspace of $P_1^{}$ than in the subspaces of $H_3^{}$ and
$L_2^-$.
The gain in entropy associated with this larger multiplicity can
energetically favor ordering at $P$, in a manner analogous to the
order-by-disorder mechanism discussed in the context of frustrated
magnetic systems \cite{Villain1980,Chubukov1993}.
The existence of almost degenerate states that are not trivially
related by symmetry is a prerequisite for this phenomenon, and the
heuristic argument made here should be confirmed by more rigorous 
Monte Carlo simulations or field theoretical studies.

Mozaffari \etal observe a sublinear relationship $\rho_{xx} \propto
T^{0.62}$ between longitudinal resistivity $\rho_{xx}$ and temperature
$T$ in ScV$_6$Sn$_6$ above the CDW transition, which they ascribe to
enhanced scattering of charge carriers in Dirac band by high density
of electrons residing at the van Hove singularities that slightly
below the Fermi level \cite{Mozaffari2023}.  However, such van Hove
singularities also exist in the electronic structure of the closely
related compound LuV$_6$Sn$_6$, but this material does not exhibit
sublinear resistivity.  Therefore, scattering between electrons in the
dispersive Dirac and flat van Hove bands are likely not the cause of
sublinear resistivity in ScV$_6$Sn$_6$.
Meanwhile, sublinear resistivity is also observed in the vanadium
kagome materials $A$V$_3$Sb$_5$ ($A$ = K, Rb, Cs) that exhibit a CDW
instability \cite{Mozaffari2023,Ortiz2021,Yin2021,Ortiz2020}.
Calculations show that nearly-degenerate distorted phases also occur
in these materials \cite{Subedi2022}.  This suggests that scattering
of charge carriers with structural fluctuations among the competing
phases may lead to sublinear resistivity.

The CDW phase of ScV$_6$Sn$_6$ has been variously refined to $R32$
\cite{Arachchige2022} and $R\overline{3}m$ \cite{Korshunov2023} space
groups from separate x-ray diffraction studies. The difficulty in
resolving the low-temperature structure might be due to freezing in of
competing phases.  There are two, three, and six possible domains
within each $xy$ plane due to the $R\overline{3}m$ $P_1^{}(a,0,0,0)$,
$P6/mmm$ $P_1^{}(a,0,a,0)$, and $P6mm$ $P_1^{}(a,b,a,b)$ order
parameters, respectively, while the $R32$ $P_1^{}(a,0,0,0) +
P_2^{}(b,0,0,0)$ order parameter can lead to four different in-plane
domains. [Multiplication of each of these numbers by three gives the
  total number of domains due to different stackings.]  Therefore,
counting the number of domains that are present in a sample can verify
the existence of multiple nearly-degenerate minima proposed here.  The
$P6mm$ $P_1^{}(a,b,a,b)$ phase lacks the inversion symmetry, and the
observation of a second harmonic generation signal would also support
the occurrence of this phase.  A more robust test would be the
presence of multiple peaks in the pair distribution function due to
different nearest-neighbor distances, which would also manifest as
different structure factor for the diffraction peaks lying in
different Brillouin zones.  NMR experiments would be another useful
probe because different order parameters due to the $P_1^{}$
instability lead to different numbers of splitting of the Wyckoff
positions.

Hu \etal have noted the existence of a nearly flat dispersion of the
soft phonon branch in the region around $H$ as it collapses near the
CDW transition, and they instead propose that fluctuations in the
reciprocal space around $H$ stabilizes the instability at $P$
\cite{Hu2023c}.  They point out that the fluctuations are suppressed at
$P$ due to the presence of a cubic term in the polynomial expansion of
the free energy, although they do not discuss the occurrence of
multiple nontrivially related minima in the energy landscape due to
the odd-order nonlinearity.  In principle, their theory is
complementary to the one proposed in the present study, although the
emphasis on the fluctuations about $H$ is hard to reconcile with the
disappearance of diffuse signals around $H$ in diffuse scattering
experiments.  
%

\section{Summary and Conclusions}

In summary, I have used first principles calculations to map out the
energy landscape of the structural distortions in ScV$_6$Sn$_6$ due to
the phonon instabilities present in its high-temperature $P6/mmm$
phase.  Consistent with previous theoretical and experimental studies,
the calculated phonon dispersions show a nondegenerate branch that is
unstable along the path $L$-$H$ and at $P$, with the instability at
$H$ being the dominant one.  I used group theoretical analysis to
enumerate all possible distortions due to the instabilities at $H$,
$L$, and $P$, and generated corresponding structures using the
calculated phonon eigenvectors.  Structural relaxations show that
distortions due to the instabilities at $H$ and $L$ have lower
calculated total energies than the ones due to the instability at $P$,
which is the wave vector where the CDW order condenses.  However, I
find that energy landscape in the submanifold defined by the order
parameter of the $P$ instability is shallower than those due to the
$H$ and $L$ instabilities.  Only two and three symmetrically-distinct
distorted structures are stable at $H$ and $L$ that are spread within
the energy ranges of 6 and 5 meV/\fu, respectively.  However, I was
able to stabilize six different structures due to the instability at
$P$ whose relative energies lie within 2 meV/\fu of each other.

The presence of a larger number of almost degenerate distorted
structures at $P$ likely provides the requisite entropic force to
cause the first-order CDW transition experimentally observed in
ScV$_6$Sn$_6$ at this wave vector via the order-by-disorder mechanism,
and the heuristic suggestion made here should be confirmed by more
rigorous theoretical studies. The energetically shallow manifold of
distortions at $P$ could also be verified experimentally, for example,
by the presence of more than two inplane domains in the
low-temperature phase or the presence of multiple peaks in the pair
distribution function due to different nearest-neighbor distances.

\section{acknowledgements}
I am grateful to Santiago Blanco-Canosa and Beno\^it Fauqu\'e for
insightful discussions.  This work was supported by GENCI-TGCC under
grant no.\ A0110913028.

\bibliography{scv6sn6-paper}

\begin{thebibliography}{44}%
\makeatletter
\providecommand \@ifxundefined [1]{%
 \@ifx{#1\undefined}
}%
\providecommand \@ifnum [1]{%
 \ifnum #1\expandafter \@firstoftwo
 \else \expandafter \@secondoftwo
 \fi
}%
\providecommand \@ifx [1]{%
 \ifx #1\expandafter \@firstoftwo
 \else \expandafter \@secondoftwo
 \fi
}%
\providecommand \natexlab [1]{#1}%
\providecommand \enquote  [1]{``#1''}%
\providecommand \bibnamefont  [1]{#1}%
\providecommand \bibfnamefont [1]{#1}%
\providecommand \citenamefont [1]{#1}%
\providecommand \href@noop [0]{\@secondoftwo}%
\providecommand \href [0]{\begingroup \@sanitize@url \@href}%
\providecommand \@href[1]{\@@startlink{#1}\@@href}%
\providecommand \@@href[1]{\endgroup#1\@@endlink}%
\providecommand \@sanitize@url [0]{\catcode `\\12\catcode `\$12\catcode
  `\&12\catcode `\#12\catcode `\^12\catcode `\_12\catcode `\%12\relax}%
\providecommand \@@startlink[1]{}%
\providecommand \@@endlink[0]{}%
\providecommand \url  [0]{\begingroup\@sanitize@url \@url }%
\providecommand \@url [1]{\endgroup\@href {#1}{\urlprefix }}%
\providecommand \urlprefix  [0]{URL }%
\providecommand \Eprint [0]{\href }%
\providecommand \doibase [0]{https://doi.org/}%
\providecommand \selectlanguage [0]{\@gobble}%
\providecommand \bibinfo  [0]{\@secondoftwo}%
\providecommand \bibfield  [0]{\@secondoftwo}%
\providecommand \translation [1]{[#1]}%
\providecommand \BibitemOpen [0]{}%
\providecommand \bibitemStop [0]{}%
\providecommand \bibitemNoStop [0]{.\EOS\space}%
\providecommand \EOS [0]{\spacefactor3000\relax}%
\providecommand \BibitemShut  [1]{\csname bibitem#1\endcsname}%
\let\auto@bib@innerbib\@empty
\bibitem [{\citenamefont {Mendels}\ and\ \citenamefont
  {Bert}(2016)}]{Mendels2016}%
  \BibitemOpen
  \bibfield  {author} {\bibinfo {author} {\bibfnamefont {P.}~\bibnamefont
  {Mendels}}\ and\ \bibinfo {author} {\bibfnamefont {F.}~\bibnamefont {Bert}},\
  }\bibfield  {title} {\bibinfo {title} {Quantum kagome frustrated
  antiferromagnets: One route to quantum spin liquids},\ }\href
  {https://doi.org/https://doi.org/10.1016/j.crhy.2015.12.001} {\bibfield
  {journal} {\bibinfo  {journal} {Comptes Rendus Physique}\ }\textbf {\bibinfo
  {volume} {17}},\ \bibinfo {pages} {455} (\bibinfo {year} {2016})}\BibitemShut
  {NoStop}%
\bibitem [{\citenamefont {Beugeling}\ \emph {et~al.}(2012)\citenamefont
  {Beugeling}, \citenamefont {Everts},\ and\ \citenamefont
  {Morais~Smith}}]{Beugeling2012}%
  \BibitemOpen
  \bibfield  {author} {\bibinfo {author} {\bibfnamefont {W.}~\bibnamefont
  {Beugeling}}, \bibinfo {author} {\bibfnamefont {J.~C.}\ \bibnamefont
  {Everts}},\ and\ \bibinfo {author} {\bibfnamefont {C.}~\bibnamefont
  {Morais~Smith}},\ }\bibfield  {title} {\bibinfo {title} {Topological phase
  transitions driven by next-nearest-neighbor hopping in two-dimensional
  lattices},\ }\href {https://doi.org/10.1103/PhysRevB.86.195129} {\bibfield
  {journal} {\bibinfo  {journal} {Phys. Rev. B}\ }\textbf {\bibinfo {volume}
  {86}},\ \bibinfo {pages} {195129} (\bibinfo {year} {2012})}\BibitemShut
  {NoStop}%
\bibitem [{\citenamefont {Meier}\ \emph {et~al.}(2020)\citenamefont {Meier},
  \citenamefont {Du}, \citenamefont {Okamoto}, \citenamefont {Mohanta},
  \citenamefont {May}, \citenamefont {McGuire}, \citenamefont {Bridges},
  \citenamefont {Samolyuk},\ and\ \citenamefont {Sales}}]{Meier2020}%
  \BibitemOpen
  \bibfield  {author} {\bibinfo {author} {\bibfnamefont {W.~R.}\ \bibnamefont
  {Meier}}, \bibinfo {author} {\bibfnamefont {M.-H.}\ \bibnamefont {Du}},
  \bibinfo {author} {\bibfnamefont {S.}~\bibnamefont {Okamoto}}, \bibinfo
  {author} {\bibfnamefont {N.}~\bibnamefont {Mohanta}}, \bibinfo {author}
  {\bibfnamefont {A.~F.}\ \bibnamefont {May}}, \bibinfo {author} {\bibfnamefont
  {M.~A.}\ \bibnamefont {McGuire}}, \bibinfo {author} {\bibfnamefont {C.~A.}\
  \bibnamefont {Bridges}}, \bibinfo {author} {\bibfnamefont {G.~D.}\
  \bibnamefont {Samolyuk}},\ and\ \bibinfo {author} {\bibfnamefont {B.~C.}\
  \bibnamefont {Sales}},\ }\bibfield  {title} {\bibinfo {title} {Flat bands in
  the cosn-type compounds},\ }\href
  {https://doi.org/10.1103/PhysRevB.102.075148} {\bibfield  {journal} {\bibinfo
   {journal} {Phys. Rev. B}\ }\textbf {\bibinfo {volume} {102}},\ \bibinfo
  {pages} {075148} (\bibinfo {year} {2020})}\BibitemShut {NoStop}%
\bibitem [{\citenamefont {Kang}\ \emph {et~al.}(2019)\citenamefont {Kang},
  \citenamefont {Ye}, \citenamefont {Fang}, \citenamefont {You}, \citenamefont
  {Levitan}, \citenamefont {Han}, \citenamefont {Facio}, \citenamefont
  {Jozwiak}, \citenamefont {Bostwick}, \citenamefont {Rotenberg}, \citenamefont
  {Chan}, \citenamefont {McDonald}, \citenamefont {Graf}, \citenamefont
  {Kaznatcheev}, \citenamefont {Vescovo}, \citenamefont {Bell}, \citenamefont
  {Kaxiras}, \citenamefont {van~den Brink}, \citenamefont {Richter},
  \citenamefont {Ghimire}, \citenamefont {Checkelsky},\ and\ \citenamefont
  {Comin}}]{Kang2019}%
  \BibitemOpen
  \bibfield  {author} {\bibinfo {author} {\bibfnamefont {M.}~\bibnamefont
  {Kang}}, \bibinfo {author} {\bibfnamefont {L.}~\bibnamefont {Ye}}, \bibinfo
  {author} {\bibfnamefont {S.}~\bibnamefont {Fang}}, \bibinfo {author}
  {\bibfnamefont {J.-S.}\ \bibnamefont {You}}, \bibinfo {author} {\bibfnamefont
  {A.}~\bibnamefont {Levitan}}, \bibinfo {author} {\bibfnamefont
  {M.}~\bibnamefont {Han}}, \bibinfo {author} {\bibfnamefont {J.~I.}\
  \bibnamefont {Facio}}, \bibinfo {author} {\bibfnamefont {C.}~\bibnamefont
  {Jozwiak}}, \bibinfo {author} {\bibfnamefont {A.}~\bibnamefont {Bostwick}},
  \bibinfo {author} {\bibfnamefont {E.}~\bibnamefont {Rotenberg}}, \bibinfo
  {author} {\bibfnamefont {M.~K.}\ \bibnamefont {Chan}}, \bibinfo {author}
  {\bibfnamefont {R.~D.}\ \bibnamefont {McDonald}}, \bibinfo {author}
  {\bibfnamefont {D.}~\bibnamefont {Graf}}, \bibinfo {author} {\bibfnamefont
  {K.}~\bibnamefont {Kaznatcheev}}, \bibinfo {author} {\bibfnamefont
  {E.}~\bibnamefont {Vescovo}}, \bibinfo {author} {\bibfnamefont {D.~C.}\
  \bibnamefont {Bell}}, \bibinfo {author} {\bibfnamefont {E.}~\bibnamefont
  {Kaxiras}}, \bibinfo {author} {\bibfnamefont {J.}~\bibnamefont {van~den
  Brink}}, \bibinfo {author} {\bibfnamefont {M.}~\bibnamefont {Richter}},
  \bibinfo {author} {\bibfnamefont {M.~P.}\ \bibnamefont {Ghimire}}, \bibinfo
  {author} {\bibfnamefont {J.~G.}\ \bibnamefont {Checkelsky}},\ and\ \bibinfo
  {author} {\bibfnamefont {R.}~\bibnamefont {Comin}},\ }\bibfield  {title}
  {\bibinfo {title} {Dirac fermions and flat bands in the ideal kagome metal
  {FeSn}},\ }\href {https://doi.org/10.1038/s41563-019-0531-0} {\bibfield
  {journal} {\bibinfo  {journal} {Nature Materials}\ }\textbf {\bibinfo
  {volume} {19}},\ \bibinfo {pages} {163} (\bibinfo {year} {2019})}\BibitemShut
  {NoStop}%
\bibitem [{\citenamefont {Liu}\ \emph {et~al.}(2020)\citenamefont {Liu},
  \citenamefont {Li}, \citenamefont {Wang}, \citenamefont {Wang}, \citenamefont
  {Wen}, \citenamefont {Jiang}, \citenamefont {Lu}, \citenamefont {Yan},
  \citenamefont {Huang}, \citenamefont {Shen}, \citenamefont {Yin},
  \citenamefont {Wang}, \citenamefont {Yin}, \citenamefont {Lei},\ and\
  \citenamefont {Wang}}]{Liu2020}%
  \BibitemOpen
  \bibfield  {author} {\bibinfo {author} {\bibfnamefont {Z.}~\bibnamefont
  {Liu}}, \bibinfo {author} {\bibfnamefont {M.}~\bibnamefont {Li}}, \bibinfo
  {author} {\bibfnamefont {Q.}~\bibnamefont {Wang}}, \bibinfo {author}
  {\bibfnamefont {G.}~\bibnamefont {Wang}}, \bibinfo {author} {\bibfnamefont
  {C.}~\bibnamefont {Wen}}, \bibinfo {author} {\bibfnamefont {K.}~\bibnamefont
  {Jiang}}, \bibinfo {author} {\bibfnamefont {X.}~\bibnamefont {Lu}}, \bibinfo
  {author} {\bibfnamefont {S.}~\bibnamefont {Yan}}, \bibinfo {author}
  {\bibfnamefont {Y.}~\bibnamefont {Huang}}, \bibinfo {author} {\bibfnamefont
  {D.}~\bibnamefont {Shen}}, \bibinfo {author} {\bibfnamefont {J.-X.}\
  \bibnamefont {Yin}}, \bibinfo {author} {\bibfnamefont {Z.}~\bibnamefont
  {Wang}}, \bibinfo {author} {\bibfnamefont {Z.}~\bibnamefont {Yin}}, \bibinfo
  {author} {\bibfnamefont {H.}~\bibnamefont {Lei}},\ and\ \bibinfo {author}
  {\bibfnamefont {S.}~\bibnamefont {Wang}},\ }\bibfield  {title} {\bibinfo
  {title} {Orbital-selective dirac fermions and extremely flat bands in
  frustrated kagome-lattice metal {CoSn}},\ }\href
  {https://doi.org/10.1038/s41467-020-17462-4} {\bibfield  {journal} {\bibinfo
  {journal} {Nature Communications}\ }\textbf {\bibinfo {volume} {11}},\
  \bibinfo {pages} {4002} (\bibinfo {year} {2020})}\BibitemShut {NoStop}%
\bibitem [{\citenamefont {Guo}\ and\ \citenamefont {Franz}(2009)}]{Guo2009}%
  \BibitemOpen
  \bibfield  {author} {\bibinfo {author} {\bibfnamefont {H.-M.}\ \bibnamefont
  {Guo}}\ and\ \bibinfo {author} {\bibfnamefont {M.}~\bibnamefont {Franz}},\
  }\bibfield  {title} {\bibinfo {title} {Topological insulator on the kagome
  lattice},\ }\href {https://doi.org/10.1103/PhysRevB.80.113102} {\bibfield
  {journal} {\bibinfo  {journal} {Phys. Rev. B}\ }\textbf {\bibinfo {volume}
  {80}},\ \bibinfo {pages} {113102} (\bibinfo {year} {2009})}\BibitemShut
  {NoStop}%
\bibitem [{\citenamefont {R\"uegg}\ and\ \citenamefont
  {Fiete}(2011)}]{Ruegg2011}%
  \BibitemOpen
  \bibfield  {author} {\bibinfo {author} {\bibfnamefont {A.}~\bibnamefont
  {R\"uegg}}\ and\ \bibinfo {author} {\bibfnamefont {G.~A.}\ \bibnamefont
  {Fiete}},\ }\bibfield  {title} {\bibinfo {title} {Fractionally charged
  topological point defects on the kagome lattice},\ }\href
  {https://doi.org/10.1103/PhysRevB.83.165118} {\bibfield  {journal} {\bibinfo
  {journal} {Phys. Rev. B}\ }\textbf {\bibinfo {volume} {83}},\ \bibinfo
  {pages} {165118} (\bibinfo {year} {2011})}\BibitemShut {NoStop}%
\bibitem [{\citenamefont {Tang}\ \emph {et~al.}(2011)\citenamefont {Tang},
  \citenamefont {Mei},\ and\ \citenamefont {Wen}}]{Tang2011}%
  \BibitemOpen
  \bibfield  {author} {\bibinfo {author} {\bibfnamefont {E.}~\bibnamefont
  {Tang}}, \bibinfo {author} {\bibfnamefont {J.-W.}\ \bibnamefont {Mei}},\ and\
  \bibinfo {author} {\bibfnamefont {X.-G.}\ \bibnamefont {Wen}},\ }\bibfield
  {title} {\bibinfo {title} {High-temperature fractional quantum hall states},\
  }\href {https://doi.org/10.1103/PhysRevLett.106.236802} {\bibfield  {journal}
  {\bibinfo  {journal} {Phys. Rev. Lett.}\ }\textbf {\bibinfo {volume} {106}},\
  \bibinfo {pages} {236802} (\bibinfo {year} {2011})}\BibitemShut {NoStop}%
\bibitem [{\citenamefont {Wang}\ \emph {et~al.}(2013)\citenamefont {Wang},
  \citenamefont {Li}, \citenamefont {Xiang},\ and\ \citenamefont
  {Wang}}]{Wang2013}%
  \BibitemOpen
  \bibfield  {author} {\bibinfo {author} {\bibfnamefont {W.-S.}\ \bibnamefont
  {Wang}}, \bibinfo {author} {\bibfnamefont {Z.-Z.}\ \bibnamefont {Li}},
  \bibinfo {author} {\bibfnamefont {Y.-Y.}\ \bibnamefont {Xiang}},\ and\
  \bibinfo {author} {\bibfnamefont {Q.-H.}\ \bibnamefont {Wang}},\ }\bibfield
  {title} {\bibinfo {title} {Competing electronic orders on kagome lattices at
  van hove filling},\ }\href {https://doi.org/10.1103/PhysRevB.87.115135}
  {\bibfield  {journal} {\bibinfo  {journal} {Phys. Rev. B}\ }\textbf {\bibinfo
  {volume} {87}},\ \bibinfo {pages} {115135} (\bibinfo {year}
  {2013})}\BibitemShut {NoStop}%
\bibitem [{\citenamefont {Kiesel}\ \emph {et~al.}(2013)\citenamefont {Kiesel},
  \citenamefont {Platt},\ and\ \citenamefont {Thomale}}]{Kiesel2013}%
  \BibitemOpen
  \bibfield  {author} {\bibinfo {author} {\bibfnamefont {M.~L.}\ \bibnamefont
  {Kiesel}}, \bibinfo {author} {\bibfnamefont {C.}~\bibnamefont {Platt}},\ and\
  \bibinfo {author} {\bibfnamefont {R.}~\bibnamefont {Thomale}},\ }\bibfield
  {title} {\bibinfo {title} {Unconventional fermi surface instabilities in the
  kagome hubbard model},\ }\href
  {https://doi.org/10.1103/PhysRevLett.110.126405} {\bibfield  {journal}
  {\bibinfo  {journal} {Phys. Rev. Lett.}\ }\textbf {\bibinfo {volume} {110}},\
  \bibinfo {pages} {126405} (\bibinfo {year} {2013})}\BibitemShut {NoStop}%
\bibitem [{\citenamefont {Mazin}\ \emph {et~al.}(2014)\citenamefont {Mazin},
  \citenamefont {Jeschke}, \citenamefont {Lechermann}, \citenamefont {Lee},
  \citenamefont {Fink}, \citenamefont {Thomale},\ and\ \citenamefont
  {Valent{\'\i}}}]{Mazin2014}%
  \BibitemOpen
  \bibfield  {author} {\bibinfo {author} {\bibfnamefont {I.}~\bibnamefont
  {Mazin}}, \bibinfo {author} {\bibfnamefont {H.~O.}\ \bibnamefont {Jeschke}},
  \bibinfo {author} {\bibfnamefont {F.}~\bibnamefont {Lechermann}}, \bibinfo
  {author} {\bibfnamefont {H.}~\bibnamefont {Lee}}, \bibinfo {author}
  {\bibfnamefont {M.}~\bibnamefont {Fink}}, \bibinfo {author} {\bibfnamefont
  {R.}~\bibnamefont {Thomale}},\ and\ \bibinfo {author} {\bibfnamefont
  {R.}~\bibnamefont {Valent{\'\i}}},\ }\bibfield  {title} {\bibinfo {title}
  {Theoretical prediction of a strongly correlated dirac metal},\ }\href
  {https://doi.org/10.1038/ncomms5261} {\bibfield  {journal} {\bibinfo
  {journal} {Nature communications}\ }\textbf {\bibinfo {volume} {5}},\
  \bibinfo {pages} {4261} (\bibinfo {year} {2014})}\BibitemShut {NoStop}%
\bibitem [{\citenamefont {Ortiz}\ \emph {et~al.}(2019)\citenamefont {Ortiz},
  \citenamefont {Gomes}, \citenamefont {Morey}, \citenamefont {Winiarski},
  \citenamefont {Bordelon}, \citenamefont {Mangum}, \citenamefont {Oswald},
  \citenamefont {Rodriguez-Rivera}, \citenamefont {Neilson}, \citenamefont
  {Wilson}, \citenamefont {Ertekin}, \citenamefont {McQueen},\ and\
  \citenamefont {Toberer}}]{Ortiz2019}%
  \BibitemOpen
  \bibfield  {author} {\bibinfo {author} {\bibfnamefont {B.~R.}\ \bibnamefont
  {Ortiz}}, \bibinfo {author} {\bibfnamefont {L.~C.}\ \bibnamefont {Gomes}},
  \bibinfo {author} {\bibfnamefont {J.~R.}\ \bibnamefont {Morey}}, \bibinfo
  {author} {\bibfnamefont {M.}~\bibnamefont {Winiarski}}, \bibinfo {author}
  {\bibfnamefont {M.}~\bibnamefont {Bordelon}}, \bibinfo {author}
  {\bibfnamefont {J.~S.}\ \bibnamefont {Mangum}}, \bibinfo {author}
  {\bibfnamefont {I.~W.~H.}\ \bibnamefont {Oswald}}, \bibinfo {author}
  {\bibfnamefont {J.~A.}\ \bibnamefont {Rodriguez-Rivera}}, \bibinfo {author}
  {\bibfnamefont {J.~R.}\ \bibnamefont {Neilson}}, \bibinfo {author}
  {\bibfnamefont {S.~D.}\ \bibnamefont {Wilson}}, \bibinfo {author}
  {\bibfnamefont {E.}~\bibnamefont {Ertekin}}, \bibinfo {author} {\bibfnamefont
  {T.~M.}\ \bibnamefont {McQueen}},\ and\ \bibinfo {author} {\bibfnamefont
  {E.~S.}\ \bibnamefont {Toberer}},\ }\bibfield  {title} {\bibinfo {title} {New
  kagome prototype materials: discovery of
  ${\mathrm{kv}}_{3}{\mathrm{sb}}_{5},{\mathrm{rbv}}_{3}{\mathrm{sb}}_{5}$, and
  ${\mathrm{csv}}_{3}{\mathrm{sb}}_{5}$},\ }\href
  {https://doi.org/10.1103/PhysRevMaterials.3.094407} {\bibfield  {journal}
  {\bibinfo  {journal} {Phys. Rev. Mater.}\ }\textbf {\bibinfo {volume} {3}},\
  \bibinfo {pages} {094407} (\bibinfo {year} {2019})}\BibitemShut {NoStop}%
\bibitem [{\citenamefont {Teng}\ \emph {et~al.}(2022)\citenamefont {Teng},
  \citenamefont {Chen}, \citenamefont {Ye}, \citenamefont {Rosenberg},
  \citenamefont {Liu}, \citenamefont {Yin}, \citenamefont {Jiang},
  \citenamefont {Oh}, \citenamefont {Hasan}, \citenamefont {Neubauer} \emph
  {et~al.}}]{Teng2022}%
  \BibitemOpen
  \bibfield  {author} {\bibinfo {author} {\bibfnamefont {X.}~\bibnamefont
  {Teng}}, \bibinfo {author} {\bibfnamefont {L.}~\bibnamefont {Chen}}, \bibinfo
  {author} {\bibfnamefont {F.}~\bibnamefont {Ye}}, \bibinfo {author}
  {\bibfnamefont {E.}~\bibnamefont {Rosenberg}}, \bibinfo {author}
  {\bibfnamefont {Z.}~\bibnamefont {Liu}}, \bibinfo {author} {\bibfnamefont
  {J.-X.}\ \bibnamefont {Yin}}, \bibinfo {author} {\bibfnamefont {Y.-X.}\
  \bibnamefont {Jiang}}, \bibinfo {author} {\bibfnamefont {J.~S.}\ \bibnamefont
  {Oh}}, \bibinfo {author} {\bibfnamefont {M.~Z.}\ \bibnamefont {Hasan}},
  \bibinfo {author} {\bibfnamefont {K.~J.}\ \bibnamefont {Neubauer}}, \emph
  {et~al.},\ }\bibfield  {title} {\bibinfo {title} {Discovery of charge density
  wave in a kagome lattice antiferromagnet},\ }\href
  {https://doi.org/10.1038/s41586-022-05034-z} {\bibfield  {journal} {\bibinfo
  {journal} {Nature}\ }\textbf {\bibinfo {volume} {609}},\ \bibinfo {pages}
  {490} (\bibinfo {year} {2022})}\BibitemShut {NoStop}%
\bibitem [{\citenamefont {Suriya~Arachchige}\ \emph {et~al.}(2022)\citenamefont
  {Suriya~Arachchige}, \citenamefont {Meier}, \citenamefont {Marshall},
  \citenamefont {Matsuoka}, \citenamefont {Xue}, \citenamefont {McGuire},
  \citenamefont {Hermann}, \citenamefont {Cao},\ and\ \citenamefont
  {Mandrus}}]{Arachchige2022}%
  \BibitemOpen
  \bibfield  {author} {\bibinfo {author} {\bibfnamefont {H.~W.}\ \bibnamefont
  {Suriya~Arachchige}}, \bibinfo {author} {\bibfnamefont {W.~R.}\ \bibnamefont
  {Meier}}, \bibinfo {author} {\bibfnamefont {M.}~\bibnamefont {Marshall}},
  \bibinfo {author} {\bibfnamefont {T.}~\bibnamefont {Matsuoka}}, \bibinfo
  {author} {\bibfnamefont {R.}~\bibnamefont {Xue}}, \bibinfo {author}
  {\bibfnamefont {M.~A.}\ \bibnamefont {McGuire}}, \bibinfo {author}
  {\bibfnamefont {R.~P.}\ \bibnamefont {Hermann}}, \bibinfo {author}
  {\bibfnamefont {H.}~\bibnamefont {Cao}},\ and\ \bibinfo {author}
  {\bibfnamefont {D.}~\bibnamefont {Mandrus}},\ }\bibfield  {title} {\bibinfo
  {title} {Charge density wave in kagome lattice intermetallic
  ${\mathrm{scv}}_{6}{\mathrm{sn}}_{6}$},\ }\href
  {https://doi.org/10.1103/PhysRevLett.129.216402} {\bibfield  {journal}
  {\bibinfo  {journal} {Phys. Rev. Lett.}\ }\textbf {\bibinfo {volume} {129}},\
  \bibinfo {pages} {216402} (\bibinfo {year} {2022})}\BibitemShut {NoStop}%
\bibitem [{\citenamefont {Tuniz}\ \emph {et~al.}(2023)\citenamefont {Tuniz},
  \citenamefont {Consiglio}, \citenamefont {Puntel}, \citenamefont {Bigi},
  \citenamefont {Enzner}, \citenamefont {Pokharel}, \citenamefont {Orgiani},
  \citenamefont {Bronsch}, \citenamefont {Parmigiani}, \citenamefont
  {Polewczyk}, \citenamefont {King}, \citenamefont {Wells}, \citenamefont
  {Zeljkovic}, \citenamefont {Carrara}, \citenamefont {Rossi}, \citenamefont
  {Fujii}, \citenamefont {Vobornik}, \citenamefont {Wilson}, \citenamefont
  {Thomale}, \citenamefont {Wehling}, \citenamefont {Sangiovanni},
  \citenamefont {Panaccione}, \citenamefont {Cilento}, \citenamefont {{Di
  Sante}},\ and\ \citenamefont {Mazzola}}]{Tuniz2023}%
  \BibitemOpen
  \bibfield  {author} {\bibinfo {author} {\bibfnamefont {M.}~\bibnamefont
  {Tuniz}}, \bibinfo {author} {\bibfnamefont {A.}~\bibnamefont {Consiglio}},
  \bibinfo {author} {\bibfnamefont {D.}~\bibnamefont {Puntel}}, \bibinfo
  {author} {\bibfnamefont {C.}~\bibnamefont {Bigi}}, \bibinfo {author}
  {\bibfnamefont {S.}~\bibnamefont {Enzner}}, \bibinfo {author} {\bibfnamefont
  {G.}~\bibnamefont {Pokharel}}, \bibinfo {author} {\bibfnamefont
  {P.}~\bibnamefont {Orgiani}}, \bibinfo {author} {\bibfnamefont
  {W.}~\bibnamefont {Bronsch}}, \bibinfo {author} {\bibfnamefont
  {F.}~\bibnamefont {Parmigiani}}, \bibinfo {author} {\bibfnamefont
  {V.}~\bibnamefont {Polewczyk}}, \bibinfo {author} {\bibfnamefont {P.~D.~C.}\
  \bibnamefont {King}}, \bibinfo {author} {\bibfnamefont {J.~W.}\ \bibnamefont
  {Wells}}, \bibinfo {author} {\bibfnamefont {I.}~\bibnamefont {Zeljkovic}},
  \bibinfo {author} {\bibfnamefont {P.}~\bibnamefont {Carrara}}, \bibinfo
  {author} {\bibfnamefont {G.}~\bibnamefont {Rossi}}, \bibinfo {author}
  {\bibfnamefont {J.}~\bibnamefont {Fujii}}, \bibinfo {author} {\bibfnamefont
  {I.}~\bibnamefont {Vobornik}}, \bibinfo {author} {\bibfnamefont {S.~D.}\
  \bibnamefont {Wilson}}, \bibinfo {author} {\bibfnamefont {R.}~\bibnamefont
  {Thomale}}, \bibinfo {author} {\bibfnamefont {T.}~\bibnamefont {Wehling}},
  \bibinfo {author} {\bibfnamefont {G.}~\bibnamefont {Sangiovanni}}, \bibinfo
  {author} {\bibfnamefont {G.}~\bibnamefont {Panaccione}}, \bibinfo {author}
  {\bibfnamefont {F.}~\bibnamefont {Cilento}}, \bibinfo {author} {\bibfnamefont
  {D.}~\bibnamefont {{Di Sante}}},\ and\ \bibinfo {author} {\bibfnamefont
  {F.}~\bibnamefont {Mazzola}},\ }\bibfield  {title} {\bibinfo {title}
  {Dynamics and resilience of the charge density wave in a bilayer kagome
  metal},\ }\Eprint {https://arxiv.org/abs/2302.10699} {arXiv:2302.10699
  [cond-mat.str-el]}  (\bibinfo {year} {2023})\BibitemShut {NoStop}%
\bibitem [{\citenamefont {Cheng}\ \emph {et~al.}(2023)\citenamefont {Cheng},
  \citenamefont {Ren}, \citenamefont {Li}, \citenamefont {Oh}, \citenamefont
  {Tan}, \citenamefont {Pokharel}, \citenamefont {DeStefano}, \citenamefont
  {Rosenberg}, \citenamefont {Guo}, \citenamefont {Zhang}, \citenamefont {Yue},
  \citenamefont {Lee}, \citenamefont {Gorovikov}, \citenamefont {Zonno},
  \citenamefont {Hashimoto}, \citenamefont {Lu}, \citenamefont {Ke},
  \citenamefont {Mazzola}, \citenamefont {Kono}, \citenamefont {Birgeneau},
  \citenamefont {Chu}, \citenamefont {Wilson}, \citenamefont {Wang},
  \citenamefont {Yan}, \citenamefont {Yi},\ and\ \citenamefont
  {Zeljkovic}}]{Cheng2023}%
  \BibitemOpen
  \bibfield  {author} {\bibinfo {author} {\bibfnamefont {S.}~\bibnamefont
  {Cheng}}, \bibinfo {author} {\bibfnamefont {Z.}~\bibnamefont {Ren}}, \bibinfo
  {author} {\bibfnamefont {H.}~\bibnamefont {Li}}, \bibinfo {author}
  {\bibfnamefont {J.}~\bibnamefont {Oh}}, \bibinfo {author} {\bibfnamefont
  {H.}~\bibnamefont {Tan}}, \bibinfo {author} {\bibfnamefont {G.}~\bibnamefont
  {Pokharel}}, \bibinfo {author} {\bibfnamefont {J.~M.}\ \bibnamefont
  {DeStefano}}, \bibinfo {author} {\bibfnamefont {E.}~\bibnamefont
  {Rosenberg}}, \bibinfo {author} {\bibfnamefont {Y.}~\bibnamefont {Guo}},
  \bibinfo {author} {\bibfnamefont {Y.}~\bibnamefont {Zhang}}, \bibinfo
  {author} {\bibfnamefont {Z.}~\bibnamefont {Yue}}, \bibinfo {author}
  {\bibfnamefont {Y.}~\bibnamefont {Lee}}, \bibinfo {author} {\bibfnamefont
  {S.}~\bibnamefont {Gorovikov}}, \bibinfo {author} {\bibfnamefont
  {M.}~\bibnamefont {Zonno}}, \bibinfo {author} {\bibfnamefont
  {M.}~\bibnamefont {Hashimoto}}, \bibinfo {author} {\bibfnamefont
  {D.}~\bibnamefont {Lu}}, \bibinfo {author} {\bibfnamefont {L.}~\bibnamefont
  {Ke}}, \bibinfo {author} {\bibfnamefont {F.}~\bibnamefont {Mazzola}},
  \bibinfo {author} {\bibfnamefont {J.}~\bibnamefont {Kono}}, \bibinfo {author}
  {\bibfnamefont {R.~J.}\ \bibnamefont {Birgeneau}}, \bibinfo {author}
  {\bibfnamefont {J.-H.}\ \bibnamefont {Chu}}, \bibinfo {author} {\bibfnamefont
  {S.~D.}\ \bibnamefont {Wilson}}, \bibinfo {author} {\bibfnamefont
  {Z.}~\bibnamefont {Wang}}, \bibinfo {author} {\bibfnamefont {B.}~\bibnamefont
  {Yan}}, \bibinfo {author} {\bibfnamefont {M.}~\bibnamefont {Yi}},\ and\
  \bibinfo {author} {\bibfnamefont {I.}~\bibnamefont {Zeljkovic}},\ }\bibfield
  {title} {\bibinfo {title} {Nanoscale visualization and spectral fingerprints
  of the charge order in scv6sn6 distinct from other kagome metals},\ }\Eprint
  {https://arxiv.org/abs/2302.12227} {arXiv:2302.12227 [cond-mat.str-el]}
  (\bibinfo {year} {2023})\BibitemShut {NoStop}%
\bibitem [{\citenamefont {Kang}\ \emph {et~al.}(2023)\citenamefont {Kang},
  \citenamefont {Li}, \citenamefont {Meier}, \citenamefont {Villanova},
  \citenamefont {Hus}, \citenamefont {Jeon}, \citenamefont {Arachchige},
  \citenamefont {Lu}, \citenamefont {Gai}, \citenamefont {Denlinger},
  \citenamefont {Moore}, \citenamefont {Yoon},\ and\ \citenamefont
  {Mandrus}}]{Kang2023}%
  \BibitemOpen
  \bibfield  {author} {\bibinfo {author} {\bibfnamefont {S.-H.}\ \bibnamefont
  {Kang}}, \bibinfo {author} {\bibfnamefont {H.}~\bibnamefont {Li}}, \bibinfo
  {author} {\bibfnamefont {W.~R.}\ \bibnamefont {Meier}}, \bibinfo {author}
  {\bibfnamefont {J.~W.}\ \bibnamefont {Villanova}}, \bibinfo {author}
  {\bibfnamefont {S.}~\bibnamefont {Hus}}, \bibinfo {author} {\bibfnamefont
  {H.}~\bibnamefont {Jeon}}, \bibinfo {author} {\bibfnamefont {H.~W.~S.}\
  \bibnamefont {Arachchige}}, \bibinfo {author} {\bibfnamefont
  {Q.}~\bibnamefont {Lu}}, \bibinfo {author} {\bibfnamefont {Z.}~\bibnamefont
  {Gai}}, \bibinfo {author} {\bibfnamefont {J.}~\bibnamefont {Denlinger}},
  \bibinfo {author} {\bibfnamefont {R.}~\bibnamefont {Moore}}, \bibinfo
  {author} {\bibfnamefont {M.}~\bibnamefont {Yoon}},\ and\ \bibinfo {author}
  {\bibfnamefont {D.}~\bibnamefont {Mandrus}},\ }\bibfield  {title} {\bibinfo
  {title} {Emergence of a new band and the lifshitz transition in kagome metal
  scv$_6$sn$_6$ with charge density wave},\ }\Eprint
  {https://arxiv.org/abs/2302.14041} {arXiv:2302.14041 [cond-mat.str-el]}
  (\bibinfo {year} {2023})\BibitemShut {NoStop}%
\bibitem [{\citenamefont {Hu}\ \emph {et~al.}(2023{\natexlab{a}})\citenamefont
  {Hu}, \citenamefont {Pi}, \citenamefont {Xu}, \citenamefont {Yue},
  \citenamefont {Wu}, \citenamefont {Liu}, \citenamefont {Zhang}, \citenamefont
  {Li}, \citenamefont {Zhou}, \citenamefont {Yuan}, \citenamefont {Wu},
  \citenamefont {Dong}, \citenamefont {Weng},\ and\ \citenamefont
  {Wang}}]{Hu2023a}%
  \BibitemOpen
  \bibfield  {author} {\bibinfo {author} {\bibfnamefont {T.}~\bibnamefont
  {Hu}}, \bibinfo {author} {\bibfnamefont {H.}~\bibnamefont {Pi}}, \bibinfo
  {author} {\bibfnamefont {S.}~\bibnamefont {Xu}}, \bibinfo {author}
  {\bibfnamefont {L.}~\bibnamefont {Yue}}, \bibinfo {author} {\bibfnamefont
  {Q.}~\bibnamefont {Wu}}, \bibinfo {author} {\bibfnamefont {Q.}~\bibnamefont
  {Liu}}, \bibinfo {author} {\bibfnamefont {S.}~\bibnamefont {Zhang}}, \bibinfo
  {author} {\bibfnamefont {R.}~\bibnamefont {Li}}, \bibinfo {author}
  {\bibfnamefont {X.}~\bibnamefont {Zhou}}, \bibinfo {author} {\bibfnamefont
  {J.}~\bibnamefont {Yuan}}, \bibinfo {author} {\bibfnamefont {D.}~\bibnamefont
  {Wu}}, \bibinfo {author} {\bibfnamefont {T.}~\bibnamefont {Dong}}, \bibinfo
  {author} {\bibfnamefont {H.}~\bibnamefont {Weng}},\ and\ \bibinfo {author}
  {\bibfnamefont {N.}~\bibnamefont {Wang}},\ }\bibfield  {title} {\bibinfo
  {title} {Optical spectroscopy and band structure calculations of the
  structural phase transition in the vanadium-based kagome metal
  ${\mathrm{scv}}_{6}{\mathrm{sn}}_{6}$},\ }\href
  {https://doi.org/10.1103/PhysRevB.107.165119} {\bibfield  {journal} {\bibinfo
   {journal} {Phys. Rev. B}\ }\textbf {\bibinfo {volume} {107}},\ \bibinfo
  {pages} {165119} (\bibinfo {year} {2023}{\natexlab{a}})}\BibitemShut
  {NoStop}%
\bibitem [{\citenamefont {Hu}\ \emph {et~al.}(2023{\natexlab{b}})\citenamefont
  {Hu}, \citenamefont {Ma}, \citenamefont {Li}, \citenamefont {Gawryluk},
  \citenamefont {Hu}, \citenamefont {Teyssier}, \citenamefont {Multian},
  \citenamefont {Yin}, \citenamefont {Jiang}, \citenamefont {Xu}, \citenamefont
  {Shin}, \citenamefont {Plokhikh}, \citenamefont {Han}, \citenamefont {Plumb},
  \citenamefont {Liu}, \citenamefont {Yin}, \citenamefont {Guguchia},
  \citenamefont {Zhao}, \citenamefont {Schnyder}, \citenamefont {Wu},
  \citenamefont {Pomjakushina}, \citenamefont {Hasan}, \citenamefont {Wang},\
  and\ \citenamefont {Shi}}]{Hu2023b}%
  \BibitemOpen
  \bibfield  {author} {\bibinfo {author} {\bibfnamefont {Y.}~\bibnamefont
  {Hu}}, \bibinfo {author} {\bibfnamefont {J.}~\bibnamefont {Ma}}, \bibinfo
  {author} {\bibfnamefont {Y.}~\bibnamefont {Li}}, \bibinfo {author}
  {\bibfnamefont {D.~J.}\ \bibnamefont {Gawryluk}}, \bibinfo {author}
  {\bibfnamefont {T.}~\bibnamefont {Hu}}, \bibinfo {author} {\bibfnamefont
  {J.}~\bibnamefont {Teyssier}}, \bibinfo {author} {\bibfnamefont
  {V.}~\bibnamefont {Multian}}, \bibinfo {author} {\bibfnamefont
  {Z.}~\bibnamefont {Yin}}, \bibinfo {author} {\bibfnamefont {Y.}~\bibnamefont
  {Jiang}}, \bibinfo {author} {\bibfnamefont {S.}~\bibnamefont {Xu}}, \bibinfo
  {author} {\bibfnamefont {S.}~\bibnamefont {Shin}}, \bibinfo {author}
  {\bibfnamefont {I.}~\bibnamefont {Plokhikh}}, \bibinfo {author}
  {\bibfnamefont {X.}~\bibnamefont {Han}}, \bibinfo {author} {\bibfnamefont
  {N.~C.}\ \bibnamefont {Plumb}}, \bibinfo {author} {\bibfnamefont
  {Y.}~\bibnamefont {Liu}}, \bibinfo {author} {\bibfnamefont {J.}~\bibnamefont
  {Yin}}, \bibinfo {author} {\bibfnamefont {Z.}~\bibnamefont {Guguchia}},
  \bibinfo {author} {\bibfnamefont {Y.}~\bibnamefont {Zhao}}, \bibinfo {author}
  {\bibfnamefont {A.~P.}\ \bibnamefont {Schnyder}}, \bibinfo {author}
  {\bibfnamefont {X.}~\bibnamefont {Wu}}, \bibinfo {author} {\bibfnamefont
  {E.}~\bibnamefont {Pomjakushina}}, \bibinfo {author} {\bibfnamefont {M.~Z.}\
  \bibnamefont {Hasan}}, \bibinfo {author} {\bibfnamefont {N.}~\bibnamefont
  {Wang}},\ and\ \bibinfo {author} {\bibfnamefont {M.}~\bibnamefont {Shi}},\
  }\bibfield  {title} {\bibinfo {title} {Phonon promoted charge density wave in
  topological kagome metal {ScV$_{6}$Sn$_{6}$}},\ }\Eprint
  {https://arxiv.org/abs/2304.06431} {arXiv:2304.06431 [cond-mat.str-el]}
  (\bibinfo {year} {2023}{\natexlab{b}})\BibitemShut {NoStop}%
\bibitem [{\citenamefont {Korshunov}\ \emph {et~al.}(2023)\citenamefont
  {Korshunov}, \citenamefont {Hu}, \citenamefont {Subires}, \citenamefont
  {Jiang}, \citenamefont {C\u{a}lug\u{a}ru}, \citenamefont {Feng},
  \citenamefont {Rajapitamahuni}, \citenamefont {Yi}, \citenamefont
  {Roychowdhury}, \citenamefont {Vergniory}, \citenamefont {Strempfer},
  \citenamefont {Shekhar}, \citenamefont {Vescovo}, \citenamefont {Chernyshov},
  \citenamefont {Said}, \citenamefont {Bosak}, \citenamefont {Felser},
  \citenamefont {Bernevig},\ and\ \citenamefont
  {Blanco-Canosa}}]{Korshunov2023}%
  \BibitemOpen
  \bibfield  {author} {\bibinfo {author} {\bibfnamefont {A.}~\bibnamefont
  {Korshunov}}, \bibinfo {author} {\bibfnamefont {H.}~\bibnamefont {Hu}},
  \bibinfo {author} {\bibfnamefont {D.}~\bibnamefont {Subires}}, \bibinfo
  {author} {\bibfnamefont {Y.}~\bibnamefont {Jiang}}, \bibinfo {author}
  {\bibfnamefont {D.}~\bibnamefont {C\u{a}lug\u{a}ru}}, \bibinfo {author}
  {\bibfnamefont {X.}~\bibnamefont {Feng}}, \bibinfo {author} {\bibfnamefont
  {A.}~\bibnamefont {Rajapitamahuni}}, \bibinfo {author} {\bibfnamefont
  {C.}~\bibnamefont {Yi}}, \bibinfo {author} {\bibfnamefont {S.}~\bibnamefont
  {Roychowdhury}}, \bibinfo {author} {\bibfnamefont {M.~G.}\ \bibnamefont
  {Vergniory}}, \bibinfo {author} {\bibfnamefont {J.}~\bibnamefont
  {Strempfer}}, \bibinfo {author} {\bibfnamefont {C.}~\bibnamefont {Shekhar}},
  \bibinfo {author} {\bibfnamefont {E.}~\bibnamefont {Vescovo}}, \bibinfo
  {author} {\bibfnamefont {D.}~\bibnamefont {Chernyshov}}, \bibinfo {author}
  {\bibfnamefont {A.~H.}\ \bibnamefont {Said}}, \bibinfo {author}
  {\bibfnamefont {A.}~\bibnamefont {Bosak}}, \bibinfo {author} {\bibfnamefont
  {C.}~\bibnamefont {Felser}}, \bibinfo {author} {\bibfnamefont {B.~A.}\
  \bibnamefont {Bernevig}},\ and\ \bibinfo {author} {\bibfnamefont
  {S.}~\bibnamefont {Blanco-Canosa}},\ }\bibfield  {title} {\bibinfo {title}
  {Softening of a flat phonon mode in the kagome {ScV$_6$Sn$_6$}},\ }\Eprint
  {https://arxiv.org/abs/2304.09173} {arXiv:2304.09173 [cond-mat.str-el]}
  (\bibinfo {year} {2023})\BibitemShut {NoStop}%
\bibitem [{\citenamefont {Lee}\ \emph {et~al.}(2023)\citenamefont {Lee},
  \citenamefont {Won}, \citenamefont {Kim}, \citenamefont {Yoo}, \citenamefont
  {Park}, \citenamefont {Denlinger}, \citenamefont {Jozwiak}, \citenamefont
  {Bostwick}, \citenamefont {Rotenberg}, \citenamefont {Comin}, \citenamefont
  {Kang},\ and\ \citenamefont {Park}}]{Lee2023}%
  \BibitemOpen
  \bibfield  {author} {\bibinfo {author} {\bibfnamefont {S.}~\bibnamefont
  {Lee}}, \bibinfo {author} {\bibfnamefont {C.}~\bibnamefont {Won}}, \bibinfo
  {author} {\bibfnamefont {J.}~\bibnamefont {Kim}}, \bibinfo {author}
  {\bibfnamefont {J.}~\bibnamefont {Yoo}}, \bibinfo {author} {\bibfnamefont
  {S.}~\bibnamefont {Park}}, \bibinfo {author} {\bibfnamefont {J.}~\bibnamefont
  {Denlinger}}, \bibinfo {author} {\bibfnamefont {C.}~\bibnamefont {Jozwiak}},
  \bibinfo {author} {\bibfnamefont {A.}~\bibnamefont {Bostwick}}, \bibinfo
  {author} {\bibfnamefont {E.}~\bibnamefont {Rotenberg}}, \bibinfo {author}
  {\bibfnamefont {R.}~\bibnamefont {Comin}}, \bibinfo {author} {\bibfnamefont
  {M.}~\bibnamefont {Kang}},\ and\ \bibinfo {author} {\bibfnamefont {J.-H.}\
  \bibnamefont {Park}},\ }\bibfield  {title} {\bibinfo {title} {Nature of
  charge density wave in kagome metal {ScV$_6$Sn$_6$}},\ }\Eprint
  {https://arxiv.org/abs/2304.11820} {arXiv:2304.11820 [cond-mat.str-el]}
  (\bibinfo {year} {2023})\BibitemShut {NoStop}%
\bibitem [{\citenamefont {{Di Sante}}\ \emph {et~al.}(2023)\citenamefont {{Di
  Sante}}, \citenamefont {Bigi}, \citenamefont {Eck}, \citenamefont {Enzner},
  \citenamefont {Consiglio}, \citenamefont {Pokharel}, \citenamefont {Carrara},
  \citenamefont {Orgiani}, \citenamefont {Polewczyk}, \citenamefont {Fujii},
  \citenamefont {King}, \citenamefont {Vobornik}, \citenamefont {Rossi},
  \citenamefont {Zeljkovic}, \citenamefont {Wilson}, \citenamefont {Thomale},
  \citenamefont {Sangiovanni}, \citenamefont {Panaccione},\ and\ \citenamefont
  {Mazzola}}]{DiSante2023}%
  \BibitemOpen
  \bibfield  {author} {\bibinfo {author} {\bibfnamefont {D.}~\bibnamefont {{Di
  Sante}}}, \bibinfo {author} {\bibfnamefont {C.}~\bibnamefont {Bigi}},
  \bibinfo {author} {\bibfnamefont {P.}~\bibnamefont {Eck}}, \bibinfo {author}
  {\bibfnamefont {S.}~\bibnamefont {Enzner}}, \bibinfo {author} {\bibfnamefont
  {A.}~\bibnamefont {Consiglio}}, \bibinfo {author} {\bibfnamefont
  {G.}~\bibnamefont {Pokharel}}, \bibinfo {author} {\bibfnamefont
  {P.}~\bibnamefont {Carrara}}, \bibinfo {author} {\bibfnamefont
  {P.}~\bibnamefont {Orgiani}}, \bibinfo {author} {\bibfnamefont
  {V.}~\bibnamefont {Polewczyk}}, \bibinfo {author} {\bibfnamefont
  {J.}~\bibnamefont {Fujii}}, \bibinfo {author} {\bibfnamefont {P.~D.~C.}\
  \bibnamefont {King}}, \bibinfo {author} {\bibfnamefont {I.}~\bibnamefont
  {Vobornik}}, \bibinfo {author} {\bibfnamefont {G.}~\bibnamefont {Rossi}},
  \bibinfo {author} {\bibfnamefont {I.}~\bibnamefont {Zeljkovic}}, \bibinfo
  {author} {\bibfnamefont {S.~D.}\ \bibnamefont {Wilson}}, \bibinfo {author}
  {\bibfnamefont {R.}~\bibnamefont {Thomale}}, \bibinfo {author} {\bibfnamefont
  {G.}~\bibnamefont {Sangiovanni}}, \bibinfo {author} {\bibfnamefont
  {G.}~\bibnamefont {Panaccione}},\ and\ \bibinfo {author} {\bibfnamefont
  {F.}~\bibnamefont {Mazzola}},\ }\bibfield  {title} {\bibinfo {title} {Flat
  band separation and robust spin berry curvature in bilayer kagome metals},\
  }\bibfield  {journal} {\bibinfo  {journal} {Nature Physics}\ }\href
  {https://doi.org/10.1038/s41567-023-02053-z} {10.1038/s41567-023-02053-z}
  (\bibinfo {year} {2023})\BibitemShut {NoStop}%
\bibitem [{\citenamefont {Cao}\ \emph {et~al.}(2023)\citenamefont {Cao},
  \citenamefont {Xu}, \citenamefont {Fukui}, \citenamefont {Manjo},
  \citenamefont {Shi}, \citenamefont {Liu}, \citenamefont {Cao},\ and\
  \citenamefont {Song}}]{Cao2023}%
  \BibitemOpen
  \bibfield  {author} {\bibinfo {author} {\bibfnamefont {S.}~\bibnamefont
  {Cao}}, \bibinfo {author} {\bibfnamefont {C.}~\bibnamefont {Xu}}, \bibinfo
  {author} {\bibfnamefont {H.}~\bibnamefont {Fukui}}, \bibinfo {author}
  {\bibfnamefont {T.}~\bibnamefont {Manjo}}, \bibinfo {author} {\bibfnamefont
  {M.}~\bibnamefont {Shi}}, \bibinfo {author} {\bibfnamefont {Y.}~\bibnamefont
  {Liu}}, \bibinfo {author} {\bibfnamefont {C.}~\bibnamefont {Cao}},\ and\
  \bibinfo {author} {\bibfnamefont {Y.}~\bibnamefont {Song}},\ }\bibfield
  {title} {\bibinfo {title} {Competing charge-density wave instabilities in the
  kagome metal {ScV$_6$Sn$_6$}},\ }\Eprint {https://arxiv.org/abs/2304.08197}
  {arXiv:2304.08197 [cond-mat.str-el]}  (\bibinfo {year} {2023})\BibitemShut
  {NoStop}%
\bibitem [{\citenamefont {Pokharel}\ \emph {et~al.}(2023)\citenamefont
  {Pokharel}, \citenamefont {Ortiz}, \citenamefont {Kautzsch}, \citenamefont
  {Alvarado}, \citenamefont {Mallayya}, \citenamefont {Wu}, \citenamefont
  {Kim}, \citenamefont {Ruff}, \citenamefont {Sarker},\ and\ \citenamefont
  {Wilson}}]{Pokharel2023}%
  \BibitemOpen
  \bibfield  {author} {\bibinfo {author} {\bibfnamefont {G.}~\bibnamefont
  {Pokharel}}, \bibinfo {author} {\bibfnamefont {B.~R.}\ \bibnamefont {Ortiz}},
  \bibinfo {author} {\bibfnamefont {L.}~\bibnamefont {Kautzsch}}, \bibinfo
  {author} {\bibfnamefont {S.~J.~G.}\ \bibnamefont {Alvarado}}, \bibinfo
  {author} {\bibfnamefont {K.}~\bibnamefont {Mallayya}}, \bibinfo {author}
  {\bibfnamefont {G.}~\bibnamefont {Wu}}, \bibinfo {author} {\bibfnamefont
  {E.-A.}\ \bibnamefont {Kim}}, \bibinfo {author} {\bibfnamefont {J.~P.~C.}\
  \bibnamefont {Ruff}}, \bibinfo {author} {\bibfnamefont {S.}~\bibnamefont
  {Sarker}},\ and\ \bibinfo {author} {\bibfnamefont {S.~D.}\ \bibnamefont
  {Wilson}},\ }\href@noop {} {\bibinfo {title} {Frustrated charge order and
  cooperative distortions in scv6sn6}} (\bibinfo {year} {2023}),\ \Eprint
  {https://arxiv.org/abs/2307.11843} {arXiv:2307.11843 [cond-mat.str-el]}
  \BibitemShut {NoStop}%
\bibitem [{\citenamefont {Tan}\ and\ \citenamefont {Yan}(2023)}]{Tan2023}%
  \BibitemOpen
  \bibfield  {author} {\bibinfo {author} {\bibfnamefont {H.}~\bibnamefont
  {Tan}}\ and\ \bibinfo {author} {\bibfnamefont {B.}~\bibnamefont {Yan}},\
  }\bibfield  {title} {\bibinfo {title} {Abundant lattice instability in kagome
  metal {ScV$_6$Sn$_6$}},\ }\Eprint {https://arxiv.org/abs/2302.07922}
  {arXiv:2302.07922 [cond-mat.mtrl-sci]}  (\bibinfo {year} {2023})\BibitemShut
  {NoStop}%
\bibitem [{\citenamefont {Giannozzi}\ \emph {et~al.}(2017)\citenamefont
  {Giannozzi}, \citenamefont {Andreussi}, \citenamefont {Brumme}, \citenamefont
  {Bunau}, \citenamefont {Nardelli}, \citenamefont {Calandra}, \citenamefont
  {Car}, \citenamefont {Cavazzoni}, \citenamefont {Ceresoli}, \citenamefont
  {Cococcioni}, \citenamefont {Colonna}, \citenamefont {Carnimeo},
  \citenamefont {Corso}, \citenamefont {de~Gironcoli}, \citenamefont {Delugas},
  \citenamefont {DiStasio}, \citenamefont {Ferretti}, \citenamefont {Floris},
  \citenamefont {Fratesi}, \citenamefont {Fugallo}, \citenamefont {Gebauer},
  \citenamefont {Gerstmann}, \citenamefont {Giustino}, \citenamefont {Gorni},
  \citenamefont {Jia}, \citenamefont {Kawamura}, \citenamefont {Ko},
  \citenamefont {Kokalj}, \citenamefont {Küçükbenli}, \citenamefont
  {Lazzeri}, \citenamefont {Marsili}, \citenamefont {Marzari}, \citenamefont
  {Mauri}, \citenamefont {Nguyen}, \citenamefont {Nguyen}, \citenamefont {de-la
  Roza}, \citenamefont {Paulatto}, \citenamefont {Poncé}, \citenamefont
  {Rocca}, \citenamefont {Sabatini}, \citenamefont {Santra}, \citenamefont
  {Schlipf}, \citenamefont {Seitsonen}, \citenamefont {Smogunov}, \citenamefont
  {Timrov}, \citenamefont {Thonhauser}, \citenamefont {Umari}, \citenamefont
  {Vast}, \citenamefont {Wu},\ and\ \citenamefont {Baroni}}]{qe}%
  \BibitemOpen
  \bibfield  {author} {\bibinfo {author} {\bibfnamefont {P.}~\bibnamefont
  {Giannozzi}}, \bibinfo {author} {\bibfnamefont {O.}~\bibnamefont
  {Andreussi}}, \bibinfo {author} {\bibfnamefont {T.}~\bibnamefont {Brumme}},
  \bibinfo {author} {\bibfnamefont {O.}~\bibnamefont {Bunau}}, \bibinfo
  {author} {\bibfnamefont {M.~B.}\ \bibnamefont {Nardelli}}, \bibinfo {author}
  {\bibfnamefont {M.}~\bibnamefont {Calandra}}, \bibinfo {author}
  {\bibfnamefont {R.}~\bibnamefont {Car}}, \bibinfo {author} {\bibfnamefont
  {C.}~\bibnamefont {Cavazzoni}}, \bibinfo {author} {\bibfnamefont
  {D.}~\bibnamefont {Ceresoli}}, \bibinfo {author} {\bibfnamefont
  {M.}~\bibnamefont {Cococcioni}}, \bibinfo {author} {\bibfnamefont
  {N.}~\bibnamefont {Colonna}}, \bibinfo {author} {\bibfnamefont
  {I.}~\bibnamefont {Carnimeo}}, \bibinfo {author} {\bibfnamefont {A.~D.}\
  \bibnamefont {Corso}}, \bibinfo {author} {\bibfnamefont {S.}~\bibnamefont
  {de~Gironcoli}}, \bibinfo {author} {\bibfnamefont {P.}~\bibnamefont
  {Delugas}}, \bibinfo {author} {\bibfnamefont {R.~A.}\ \bibnamefont
  {DiStasio}}, \bibinfo {author} {\bibfnamefont {A.}~\bibnamefont {Ferretti}},
  \bibinfo {author} {\bibfnamefont {A.}~\bibnamefont {Floris}}, \bibinfo
  {author} {\bibfnamefont {G.}~\bibnamefont {Fratesi}}, \bibinfo {author}
  {\bibfnamefont {G.}~\bibnamefont {Fugallo}}, \bibinfo {author} {\bibfnamefont
  {R.}~\bibnamefont {Gebauer}}, \bibinfo {author} {\bibfnamefont
  {U.}~\bibnamefont {Gerstmann}}, \bibinfo {author} {\bibfnamefont
  {F.}~\bibnamefont {Giustino}}, \bibinfo {author} {\bibfnamefont
  {T.}~\bibnamefont {Gorni}}, \bibinfo {author} {\bibfnamefont
  {J.}~\bibnamefont {Jia}}, \bibinfo {author} {\bibfnamefont {M.}~\bibnamefont
  {Kawamura}}, \bibinfo {author} {\bibfnamefont {H.-Y.}\ \bibnamefont {Ko}},
  \bibinfo {author} {\bibfnamefont {A.}~\bibnamefont {Kokalj}}, \bibinfo
  {author} {\bibfnamefont {E.}~\bibnamefont {Küçükbenli}}, \bibinfo {author}
  {\bibfnamefont {M.}~\bibnamefont {Lazzeri}}, \bibinfo {author} {\bibfnamefont
  {M.}~\bibnamefont {Marsili}}, \bibinfo {author} {\bibfnamefont
  {N.}~\bibnamefont {Marzari}}, \bibinfo {author} {\bibfnamefont
  {F.}~\bibnamefont {Mauri}}, \bibinfo {author} {\bibfnamefont {N.~L.}\
  \bibnamefont {Nguyen}}, \bibinfo {author} {\bibfnamefont {H.-V.}\
  \bibnamefont {Nguyen}}, \bibinfo {author} {\bibfnamefont {A.~O.}\
  \bibnamefont {de-la Roza}}, \bibinfo {author} {\bibfnamefont
  {L.}~\bibnamefont {Paulatto}}, \bibinfo {author} {\bibfnamefont
  {S.}~\bibnamefont {Poncé}}, \bibinfo {author} {\bibfnamefont
  {D.}~\bibnamefont {Rocca}}, \bibinfo {author} {\bibfnamefont
  {R.}~\bibnamefont {Sabatini}}, \bibinfo {author} {\bibfnamefont
  {B.}~\bibnamefont {Santra}}, \bibinfo {author} {\bibfnamefont
  {M.}~\bibnamefont {Schlipf}}, \bibinfo {author} {\bibfnamefont {A.~P.}\
  \bibnamefont {Seitsonen}}, \bibinfo {author} {\bibfnamefont {A.}~\bibnamefont
  {Smogunov}}, \bibinfo {author} {\bibfnamefont {I.}~\bibnamefont {Timrov}},
  \bibinfo {author} {\bibfnamefont {T.}~\bibnamefont {Thonhauser}}, \bibinfo
  {author} {\bibfnamefont {P.}~\bibnamefont {Umari}}, \bibinfo {author}
  {\bibfnamefont {N.}~\bibnamefont {Vast}}, \bibinfo {author} {\bibfnamefont
  {X.}~\bibnamefont {Wu}},\ and\ \bibinfo {author} {\bibfnamefont
  {S.}~\bibnamefont {Baroni}},\ }\bibfield  {title} {\bibinfo {title} {Advanced
  capabilities for materials modelling with quantum espresso},\ }\href
  {https://doi.org/10.1088/1361-648X/aa8f79} {\bibfield  {journal} {\bibinfo
  {journal} {Journal of Physics: Condensed Matter}\ }\textbf {\bibinfo {volume}
  {29}},\ \bibinfo {pages} {465901} (\bibinfo {year} {2017})}\BibitemShut
  {NoStop}%
\bibitem [{\citenamefont {J.~Klime\v{s}}(2009)}]{optb88}%
  \BibitemOpen
  \bibfield  {author} {\bibinfo {author} {\bibfnamefont {A.~M.}\ \bibnamefont
  {J.~Klime\v{s}}, \bibfnamefont {D.~R.~Bowler}},\ }\bibfield  {title}
  {\bibinfo {title} {Chemical accuracy for the van der waals density
  functional},\ }\href {https://doi.org/10.1088/0953-8984/22/2/022201}
  {\bibfield  {journal} {\bibinfo  {journal} {Journal of Physics: Condensed
  Matter}\ }\textbf {\bibinfo {volume} {22}},\ \bibinfo {pages} {022201}
  (\bibinfo {year} {2009})}\BibitemShut {NoStop}%
\bibitem [{\citenamefont {Dal~Corso}(2014)}]{pslib}%
  \BibitemOpen
  \bibfield  {author} {\bibinfo {author} {\bibfnamefont {A.}~\bibnamefont
  {Dal~Corso}},\ }\bibfield  {title} {\bibinfo {title} {Pseudopotentials
  periodic table: From {H} to {Pu}},\ }\href@noop {} {\bibfield  {journal}
  {\bibinfo  {journal} {Computational Materials Science}\ }\textbf {\bibinfo
  {volume} {95}},\ \bibinfo {pages} {337} (\bibinfo {year} {2014})}\BibitemShut
  {NoStop}%
\bibitem [{\citenamefont {Baroni}\ \emph {et~al.}(2001)\citenamefont {Baroni},
  \citenamefont {de~Gironcoli}, \citenamefont {Dal~Corso},\ and\ \citenamefont
  {Giannozzi}}]{dfpt}%
  \BibitemOpen
  \bibfield  {author} {\bibinfo {author} {\bibfnamefont {S.}~\bibnamefont
  {Baroni}}, \bibinfo {author} {\bibfnamefont {S.}~\bibnamefont
  {de~Gironcoli}}, \bibinfo {author} {\bibfnamefont {A.}~\bibnamefont
  {Dal~Corso}},\ and\ \bibinfo {author} {\bibfnamefont {P.}~\bibnamefont
  {Giannozzi}},\ }\bibfield  {title} {\bibinfo {title} {Phonons and related
  crystal properties from density-functional perturbation theory},\ }\href
  {https://doi.org/10.1103/RevModPhys.73.515} {\bibfield  {journal} {\bibinfo
  {journal} {Rev. Mod. Phys.}\ }\textbf {\bibinfo {volume} {73}},\ \bibinfo
  {pages} {515} (\bibinfo {year} {2001})}\BibitemShut {NoStop}%
\bibitem [{\citenamefont {Stokes}\ \emph {et~al.}()\citenamefont {Stokes},
  \citenamefont {Campbell},\ and\ \citenamefont {Hatch}}]{isotropy}%
  \BibitemOpen
  \bibfield  {author} {\bibinfo {author} {\bibfnamefont {H.~T.}\ \bibnamefont
  {Stokes}}, \bibinfo {author} {\bibfnamefont {B.~J.}\ \bibnamefont
  {Campbell}},\ and\ \bibinfo {author} {\bibfnamefont {D.~M.}\ \bibnamefont
  {Hatch}},\ }\href@noop {} {\bibinfo {title} {{\sc isotropy} software
  suite}},\ \bibinfo {howpublished} {\url{iso.byu.edu}}\BibitemShut {NoStop}%
\bibitem [{\citenamefont {Stokes}\ and\ \citenamefont {Hatch}(2005)}]{findsym}%
  \BibitemOpen
  \bibfield  {author} {\bibinfo {author} {\bibfnamefont {H.~T.}\ \bibnamefont
  {Stokes}}\ and\ \bibinfo {author} {\bibfnamefont {D.~M.}\ \bibnamefont
  {Hatch}},\ }\bibfield  {title} {\bibinfo {title} {{\sc findsym}: program for
  identifying the space-group symmetry of a crystal},\ }\href
  {https://doi.org/10.1107/S0021889804031528} {\bibfield  {journal} {\bibinfo
  {journal} {Journal of Applied Crystallography}\ }\textbf {\bibinfo {volume}
  {38}},\ \bibinfo {pages} {237} (\bibinfo {year} {2005})}\BibitemShut
  {NoStop}%
\bibitem [{\citenamefont {Togo}\ and\ \citenamefont {Tanaka}(2018)}]{spglib}%
  \BibitemOpen
  \bibfield  {author} {\bibinfo {author} {\bibfnamefont {A.}~\bibnamefont
  {Togo}}\ and\ \bibinfo {author} {\bibfnamefont {I.}~\bibnamefont {Tanaka}},\
  }\bibfield  {title} {\bibinfo {title} {{\sc spglib}: a software library for
  crystal symmetry search},\ }\Eprint {https://arxiv.org/abs/1808.01590}
  {arXiv:1808.01590 [cond-mat.mtrl-sci]}  (\bibinfo {year} {2018})\BibitemShut
  {NoStop}%
\bibitem [{\citenamefont {Orobengoa}\ \emph {et~al.}(2009)\citenamefont
  {Orobengoa}, \citenamefont {Capillas}, \citenamefont {Aroyo},\ and\
  \citenamefont {Perez-Mato}}]{ampli}%
  \BibitemOpen
  \bibfield  {author} {\bibinfo {author} {\bibfnamefont {D.}~\bibnamefont
  {Orobengoa}}, \bibinfo {author} {\bibfnamefont {C.}~\bibnamefont {Capillas}},
  \bibinfo {author} {\bibfnamefont {M.~I.}\ \bibnamefont {Aroyo}},\ and\
  \bibinfo {author} {\bibfnamefont {J.~M.}\ \bibnamefont {Perez-Mato}},\
  }\bibfield  {title} {\bibinfo {title} {{\sc amplimodes}: symmetry-mode
  analysis on the bilbao crystallographic server},\ }\href
  {https://doi.org/10.1107/S0021889809028064} {\bibfield  {journal} {\bibinfo
  {journal} {Journal of Applied Crystallography}\ }\textbf {\bibinfo {volume}
  {42}},\ \bibinfo {pages} {820} (\bibinfo {year} {2009})}\BibitemShut
  {NoStop}%
\bibitem [{\citenamefont {Hatch}\ and\ \citenamefont {Stokes}(2003)}]{inv}%
  \BibitemOpen
  \bibfield  {author} {\bibinfo {author} {\bibfnamefont {D.~M.}\ \bibnamefont
  {Hatch}}\ and\ \bibinfo {author} {\bibfnamefont {H.~T.}\ \bibnamefont
  {Stokes}},\ }\bibfield  {title} {\bibinfo {title} {{{\sc invariants}: program
  for obtaining a list of invariant polynomials of the order-parameter
  components associated with irreducible representations of a space group}},\
  }\href {https://doi.org/10.1107/S0021889803005946} {\bibfield  {journal}
  {\bibinfo  {journal} {Journal of Applied Crystallography}\ }\textbf {\bibinfo
  {volume} {36}},\ \bibinfo {pages} {951} (\bibinfo {year} {2003})}\BibitemShut
  {NoStop}%
\bibitem [{sm()}]{sm}%
  \BibitemOpen
  \href@noop {} {}\bibinfo {note} {See Supplemental Material for full structual
  information of all the phases discussed in the paper.}\BibitemShut {Stop}%
\bibitem [{\citenamefont {Gu}\ \emph {et~al.}(2023)\citenamefont {Gu},
  \citenamefont {Ritz}, \citenamefont {Meier}, \citenamefont {Blockmon},
  \citenamefont {Smith}, \citenamefont {Madhogaria}, \citenamefont {Mozaffari},
  \citenamefont {Mandrus}, \citenamefont {Birol},\ and\ \citenamefont
  {Musfeldt}}]{Gu2023}%
  \BibitemOpen
  \bibfield  {author} {\bibinfo {author} {\bibfnamefont {Y.}~\bibnamefont
  {Gu}}, \bibinfo {author} {\bibfnamefont {E.}~\bibnamefont {Ritz}}, \bibinfo
  {author} {\bibfnamefont {W.~R.}\ \bibnamefont {Meier}}, \bibinfo {author}
  {\bibfnamefont {A.}~\bibnamefont {Blockmon}}, \bibinfo {author}
  {\bibfnamefont {K.}~\bibnamefont {Smith}}, \bibinfo {author} {\bibfnamefont
  {R.~P.}\ \bibnamefont {Madhogaria}}, \bibinfo {author} {\bibfnamefont
  {S.}~\bibnamefont {Mozaffari}}, \bibinfo {author} {\bibfnamefont
  {D.}~\bibnamefont {Mandrus}}, \bibinfo {author} {\bibfnamefont
  {T.}~\bibnamefont {Birol}},\ and\ \bibinfo {author} {\bibfnamefont {J.~L.}\
  \bibnamefont {Musfeldt}},\ }\bibfield  {title} {\bibinfo {title} {Origin and
  stability of the charge density wave in {ScV$_6$Sn$_6$}},\ }\Eprint
  {https://arxiv.org/abs/2305.01086} {arXiv:2305.01086 [cond-mat.str-el]}
  (\bibinfo {year} {2023})\BibitemShut {NoStop}%
\bibitem [{\citenamefont {Villain}\ \emph {et~al.}(1980)\citenamefont
  {Villain}, \citenamefont {Bidaux}, \citenamefont {Carton},\ and\
  \citenamefont {Conte}}]{Villain1980}%
  \BibitemOpen
  \bibfield  {author} {\bibinfo {author} {\bibfnamefont {J.}~\bibnamefont
  {Villain}}, \bibinfo {author} {\bibfnamefont {R.}~\bibnamefont {Bidaux}},
  \bibinfo {author} {\bibfnamefont {J.-P.}\ \bibnamefont {Carton}},\ and\
  \bibinfo {author} {\bibfnamefont {R.}~\bibnamefont {Conte}},\ }\bibfield
  {title} {\bibinfo {title} {Order as an effect of disorder},\ }\href
  {https://doi.org/10.1051/jphys:0198000410110126300} {\bibfield  {journal}
  {\bibinfo  {journal} {J. Phys. France}\ }\textbf {\bibinfo {volume} {41}},\
  \bibinfo {pages} {1263} (\bibinfo {year} {1980})}\BibitemShut {NoStop}%
\bibitem [{\citenamefont {Chubukov}(1993)}]{Chubukov1993}%
  \BibitemOpen
  \bibfield  {author} {\bibinfo {author} {\bibfnamefont {A.}~\bibnamefont
  {Chubukov}},\ }\bibfield  {title} {\bibinfo {title} {Order from disorder in a
  kagome antiferromagnet},\ }\href@noop {} {\bibfield  {journal} {\bibinfo
  {journal} {Journal of Applied Physics}\ }\textbf {\bibinfo {volume} {73}},\
  \bibinfo {pages} {5639} (\bibinfo {year} {1993})}\BibitemShut {NoStop}%
\bibitem [{\citenamefont {Mozaffari}\ \emph {et~al.}(2023)\citenamefont
  {Mozaffari}, \citenamefont {Meier}, \citenamefont {Madhogaria}, \citenamefont
  {Kang}, \citenamefont {Villanova}, \citenamefont {Arachchige}, \citenamefont
  {Zheng}, \citenamefont {Zhu}, \citenamefont {Chen}, \citenamefont {Jenkins},
  \citenamefont {Zhang}, \citenamefont {Chan}, \citenamefont {Li},
  \citenamefont {Yoon}, \citenamefont {Zhang},\ and\ \citenamefont
  {Mandrus}}]{Mozaffari2023}%
  \BibitemOpen
  \bibfield  {author} {\bibinfo {author} {\bibfnamefont {S.}~\bibnamefont
  {Mozaffari}}, \bibinfo {author} {\bibfnamefont {W.~R.}\ \bibnamefont
  {Meier}}, \bibinfo {author} {\bibfnamefont {R.~P.}\ \bibnamefont
  {Madhogaria}}, \bibinfo {author} {\bibfnamefont {S.-H.}\ \bibnamefont
  {Kang}}, \bibinfo {author} {\bibfnamefont {J.~W.}\ \bibnamefont {Villanova}},
  \bibinfo {author} {\bibfnamefont {H.~W.~S.}\ \bibnamefont {Arachchige}},
  \bibinfo {author} {\bibfnamefont {G.}~\bibnamefont {Zheng}}, \bibinfo
  {author} {\bibfnamefont {Y.}~\bibnamefont {Zhu}}, \bibinfo {author}
  {\bibfnamefont {K.-W.}\ \bibnamefont {Chen}}, \bibinfo {author}
  {\bibfnamefont {K.}~\bibnamefont {Jenkins}}, \bibinfo {author} {\bibfnamefont
  {D.}~\bibnamefont {Zhang}}, \bibinfo {author} {\bibfnamefont
  {A.}~\bibnamefont {Chan}}, \bibinfo {author} {\bibfnamefont {L.}~\bibnamefont
  {Li}}, \bibinfo {author} {\bibfnamefont {M.}~\bibnamefont {Yoon}}, \bibinfo
  {author} {\bibfnamefont {Y.}~\bibnamefont {Zhang}},\ and\ \bibinfo {author}
  {\bibfnamefont {D.~G.}\ \bibnamefont {Mandrus}},\ }\href@noop {} {\bibinfo
  {title} {Universal sublinear resistivity in vanadium kagome materials hosting
  charge density waves}} (\bibinfo {year} {2023}),\ \Eprint
  {https://arxiv.org/abs/2305.02393} {arXiv:2305.02393 [cond-mat.str-el]}
  \BibitemShut {NoStop}%
\bibitem [{\citenamefont {Ortiz}\ \emph {et~al.}(2021)\citenamefont {Ortiz},
  \citenamefont {Sarte}, \citenamefont {Kenney}, \citenamefont {Graf},
  \citenamefont {Teicher}, \citenamefont {Seshadri},\ and\ \citenamefont
  {Wilson}}]{Ortiz2021}%
  \BibitemOpen
  \bibfield  {author} {\bibinfo {author} {\bibfnamefont {B.~R.}\ \bibnamefont
  {Ortiz}}, \bibinfo {author} {\bibfnamefont {P.~M.}\ \bibnamefont {Sarte}},
  \bibinfo {author} {\bibfnamefont {E.~M.}\ \bibnamefont {Kenney}}, \bibinfo
  {author} {\bibfnamefont {M.~J.}\ \bibnamefont {Graf}}, \bibinfo {author}
  {\bibfnamefont {S.~M.~L.}\ \bibnamefont {Teicher}}, \bibinfo {author}
  {\bibfnamefont {R.}~\bibnamefont {Seshadri}},\ and\ \bibinfo {author}
  {\bibfnamefont {S.~D.}\ \bibnamefont {Wilson}},\ }\bibfield  {title}
  {\bibinfo {title} {Superconductivity in the {${\mathbb{Z}}_{2}$} kagome metal
  {${\mathrm{KV}}_{3}{\mathrm{Sb}}_{5}$}},\ }\href
  {https://doi.org/10.1103/PhysRevMaterials.5.034801} {\bibfield  {journal}
  {\bibinfo  {journal} {Phys. Rev. Mater.}\ }\textbf {\bibinfo {volume} {5}},\
  \bibinfo {pages} {034801} (\bibinfo {year} {2021})}\BibitemShut {NoStop}%
\bibitem [{\citenamefont {Yin}\ \emph {et~al.}(2021)\citenamefont {Yin},
  \citenamefont {Tu}, \citenamefont {Gong}, \citenamefont {Fu}, \citenamefont
  {Yan},\ and\ \citenamefont {Lei}}]{Yin2021}%
  \BibitemOpen
  \bibfield  {author} {\bibinfo {author} {\bibfnamefont {Q.}~\bibnamefont
  {Yin}}, \bibinfo {author} {\bibfnamefont {Z.}~\bibnamefont {Tu}}, \bibinfo
  {author} {\bibfnamefont {C.}~\bibnamefont {Gong}}, \bibinfo {author}
  {\bibfnamefont {Y.}~\bibnamefont {Fu}}, \bibinfo {author} {\bibfnamefont
  {S.}~\bibnamefont {Yan}},\ and\ \bibinfo {author} {\bibfnamefont
  {H.}~\bibnamefont {Lei}},\ }\bibfield  {title} {\bibinfo {title}
  {Superconductivity and normal-state properties of kagome metal rbv3sb5 single
  crystals},\ }\href {https://doi.org/10.1088/0256-307X/38/3/037403} {\bibfield
   {journal} {\bibinfo  {journal} {Chinese Physics Letters}\ }\textbf {\bibinfo
  {volume} {38}},\ \bibinfo {pages} {037403} (\bibinfo {year}
  {2021})}\BibitemShut {NoStop}%
\bibitem [{\citenamefont {Ortiz}\ \emph {et~al.}(2020)\citenamefont {Ortiz},
  \citenamefont {Teicher}, \citenamefont {Hu}, \citenamefont {Zuo},
  \citenamefont {Sarte}, \citenamefont {Schueller}, \citenamefont {Abeykoon},
  \citenamefont {Krogstad}, \citenamefont {Rosenkranz}, \citenamefont {Osborn},
  \citenamefont {Seshadri}, \citenamefont {Balents}, \citenamefont {He},\ and\
  \citenamefont {Wilson}}]{Ortiz2020}%
  \BibitemOpen
  \bibfield  {author} {\bibinfo {author} {\bibfnamefont {B.~R.}\ \bibnamefont
  {Ortiz}}, \bibinfo {author} {\bibfnamefont {S.~M.~L.}\ \bibnamefont
  {Teicher}}, \bibinfo {author} {\bibfnamefont {Y.}~\bibnamefont {Hu}},
  \bibinfo {author} {\bibfnamefont {J.~L.}\ \bibnamefont {Zuo}}, \bibinfo
  {author} {\bibfnamefont {P.~M.}\ \bibnamefont {Sarte}}, \bibinfo {author}
  {\bibfnamefont {E.~C.}\ \bibnamefont {Schueller}}, \bibinfo {author}
  {\bibfnamefont {A.~M.~M.}\ \bibnamefont {Abeykoon}}, \bibinfo {author}
  {\bibfnamefont {M.~J.}\ \bibnamefont {Krogstad}}, \bibinfo {author}
  {\bibfnamefont {S.}~\bibnamefont {Rosenkranz}}, \bibinfo {author}
  {\bibfnamefont {R.}~\bibnamefont {Osborn}}, \bibinfo {author} {\bibfnamefont
  {R.}~\bibnamefont {Seshadri}}, \bibinfo {author} {\bibfnamefont
  {L.}~\bibnamefont {Balents}}, \bibinfo {author} {\bibfnamefont
  {J.}~\bibnamefont {He}},\ and\ \bibinfo {author} {\bibfnamefont {S.~D.}\
  \bibnamefont {Wilson}},\ }\bibfield  {title} {\bibinfo {title}
  {$\mathrm{Cs}{\mathrm{v}}_{3}{\mathrm{sb}}_{5}$: A ${\mathbb{z}}_{2}$
  topological kagome metal with a superconducting ground state},\ }\href
  {https://doi.org/10.1103/PhysRevLett.125.247002} {\bibfield  {journal}
  {\bibinfo  {journal} {Phys. Rev. Lett.}\ }\textbf {\bibinfo {volume} {125}},\
  \bibinfo {pages} {247002} (\bibinfo {year} {2020})}\BibitemShut {NoStop}%
\bibitem [{\citenamefont {Subedi}(2022)}]{Subedi2022}%
  \BibitemOpen
  \bibfield  {author} {\bibinfo {author} {\bibfnamefont {A.}~\bibnamefont
  {Subedi}},\ }\bibfield  {title} {\bibinfo {title}
  {Hexagonal-to-base-centered-orthorhombic {$4Q$} charge density wave order in
  kagome metals {${\mathrm{KV}}_{3}{\mathrm{Sb}}_{5},$
  ${\mathrm{RbV}}_{3}{\mathrm{Sb}}_{5},$} and
  {${\mathrm{CsV}}_{3}{\mathrm{Sb}}_{5}$}},\ }\href
  {https://doi.org/10.1103/PhysRevMaterials.6.015001} {\bibfield  {journal}
  {\bibinfo  {journal} {Phys. Rev. Mater.}\ }\textbf {\bibinfo {volume} {6}},\
  \bibinfo {pages} {015001} (\bibinfo {year} {2022})}\BibitemShut {NoStop}%
\bibitem [{\citenamefont {Hu}\ \emph {et~al.}(2023{\natexlab{c}})\citenamefont
  {Hu}, \citenamefont {Jiang}, \citenamefont {C\u{a}lug\u{a}ru}, \citenamefont
  {Feng}, \citenamefont {Subires}, \citenamefont {Vergniory}, \citenamefont
  {Felser}, \citenamefont {Blanco-Canosa},\ and\ \citenamefont
  {Bernevig}}]{Hu2023c}%
  \BibitemOpen
  \bibfield  {author} {\bibinfo {author} {\bibfnamefont {H.}~\bibnamefont
  {Hu}}, \bibinfo {author} {\bibfnamefont {Y.}~\bibnamefont {Jiang}}, \bibinfo
  {author} {\bibfnamefont {D.}~\bibnamefont {C\u{a}lug\u{a}ru}}, \bibinfo
  {author} {\bibfnamefont {X.}~\bibnamefont {Feng}}, \bibinfo {author}
  {\bibfnamefont {D.}~\bibnamefont {Subires}}, \bibinfo {author} {\bibfnamefont
  {M.~G.}\ \bibnamefont {Vergniory}}, \bibinfo {author} {\bibfnamefont
  {C.}~\bibnamefont {Felser}}, \bibinfo {author} {\bibfnamefont
  {S.}~\bibnamefont {Blanco-Canosa}},\ and\ \bibinfo {author} {\bibfnamefont
  {B.~A.}\ \bibnamefont {Bernevig}},\ }\bibfield  {title} {\bibinfo {title}
  {Kagome materials {I}: Sg 191, {ScV$_6$Sn$_6$}. {F}lat {P}honon {S}oft
  {M}odes and {U}nconventional {CDW} {F}ormation: {M}icroscopic and {E}ffective
  {T}heory},\ }\Eprint {https://arxiv.org/abs/2305.15469} {arXiv:2305.15469
  [cond-mat.str-el]}  (\bibinfo {year} {2023}{\natexlab{c}})\BibitemShut
  {NoStop}%
\end{thebibliography}%

\end{document}